\documentclass{ar2e}

\usepackage{ulem}  
\usepackage{ARAstroBib}

\begin{document}
\input epsf.tex    
\input epsf.def   
\input psfig.sty

\def\araa   {\hbox{Annu. Rev. Astro. Astrophys.}}
\def\aapr   {\hbox{AAPR?}}
\def\apj     {\hbox{Astrophysical Journal}}
\def\aj       {\hbox{Astronomical Journal}}
\def\aap    {\hbox{Astronomy \& Astrophysics}}
\def\apss   {\hbox{Astronomy \& Astrophysics Supp. Ser.}}
\def\apjs   {\hbox{Astrophysical Journal Supplement}}
\def\apjl    {\hbox{Astrophysical Journal Letters}}
\def\mnras  {\hbox{Monthly Notices of the R.A.S.}}
\def\memsai  {\hbox{Mem.S.A.It.}}
\def\na       {\hbox{New Astronomy}}
\def\nat       {\hbox{Nature}}
\def\pasa    {\hbox{PASA}}
\def\pasj     {\hbox{PASJ}}
\def\pasp    {\hbox{PASP}}
\def\mic     {\hbox{$\mu m$}}
\def\kms   {\hbox{km s$^{-1}$}}
\def\Mdot  {\hbox{$M_\odot$}}

\jname{Annu. Rev. Astro. Astrophys.}
\jyear{2012}
\jvol{1}
\ARinfo{1056-8700/97/0610-00}


\title{Galactic Stellar Populations in the Era of SDSS and Other Large Surveys}

\markboth{Ivezi\'{c}, Beers \& Juri\'{c}}{Galactic Stellar Populations in the Era of SDSS and Other Large Surveys}

\author{
\v{Z}eljko Ivezi\'{c}
\affiliation{Department of Astronomy, University of Washington, Box 351580, Seattle, WA 98195}
Timothy C. Beers
\affiliation{National Optical Astronomy Observatory, Tucson, AZ, 85719, Department of Physics \& Astronomy 
and JINA: Joint Institute for Nuclear Astrophysics, Michigan State University, East Lansing, MI 48824}
Mario Juri\'{c}
\affiliation{Harvard-Smithsonian Center for Astrophysics, 60 Garden Street, Cambridge, MA
02138; Hubble Fellow}}


\begin{abstract}

Studies of stellar populations, understood to mean collections of stars with
common spatial, kinematic, chemical, and/or age distributions, have been
reinvigorated during the last decade by the advent of large-area sky surveys
such as SDSS, 2MASS, RAVE, and others. We review recent analyses of these data
that, together with theoretical and modeling advances, are revolutionizing our
understanding of the nature of the Milky Way, and galaxy formation and evolution
in general. The formation of galaxies like the Milky Way was long thought to be
a steady process leading to a smooth distribution of stars. However, the
abundance of substructure in the multi-dimensional space of various observables,
such as position, kinematics, and metallicity, is by now proven beyond doubt,
and demonstrates the importance of mergers in the growth of galaxies. Unlike
smooth models that involve simple components, the new data reviewed here clearly
exhibit many irregular structures, such as the Sagittarius dwarf tidal stream and
the Virgo and Pisces overdensities in the halo, and the Monoceros stream closer
to the Galactic plane. These recent developments have made it clear that the
Milky Way is a complex and dynamic structure, one that is still being shaped
by the merging of neighboring smaller galaxies. We also briefly discuss the next
generation of wide-field sky surveys, such as SkyMapper, Pan-STARRS, Gaia, and
LSST, which will improve measurement precision manyfold, and comprise billions
of individual stars. The ultimate goal, development of a coherent and detailed
story of the assembly and evolutionary history of the Milky Way and other large
spirals like it, now appears well within reach. 

\end{abstract}
\maketitle

\section{INTRODUCTION}

\subsection{The Big Picture: Structure Formation and\\ Near-field Cosmology}

The current cosmological paradigm states that the Universe had its beginning in
the Big Bang. Galaxies, the fundamental (luminous) building blocks of the
Universe, began forming relatively soon after this event (no more than a few
Gyr). A major objective of modern astrophysics is to understand when and how
galaxies formed, and how they have evolved since. Our own galaxy, the Milky Way,
provides a unique opportunity to study a galaxy in exquisite detail, by
measuring and analyzing the properties of large samples of individual stars.
Characterization of the stellar populations of the Milky Way provides { clues
about galaxy formation and evolution that cannot be extracted from observations
of distant galaxies alone.} Indeed, it is not possible to tell a coherent story
of the formation of the first stars and galaxies {without} understanding the
nature of the stellar populations of the Milky Way.

In the canonical model of Milky Way formation \citep*{ELS62} the Galaxy began
with a relatively rapid ($\sim 10^8$ yr) radial collapse of the initial
protogalactic cloud, followed by an equally rapid settling of gas into a
rotating disk. The ELS scenario readily explained the origin and general
structural, kinematic, and metallicity correlations of observationally
identified populations of field stars, and implied a smooth distribution of
stars observable today. The predictions of the ELS scenario were quantified by
the \cite{BS1980} and \cite{GWK1989} models, and reviewed in detail by, e.g.,
\cite{Majewski1993}. In these smooth models, the Milky Way in the relatively
nearby region (within $\sim 5$ kpc) is usually modeled by three discrete
components described by fairly simple analytic expressions: the thin disk,
the thick disk, and the halo. The other known components, the bulge and the bar,
are not expected to directly contribute to the stellar populations in this local
region.

However, for some time, starting with the pioneering work of \citet{SearleZinn},
and culminating with recent discoveries of complex substructure in the
distribution of the Milky Way's stars, this standard view has experienced
difficulties. Unlike the smooth models with simple components that have been
used on local scales, new data on larger scales indicate the presence of much
more irregular structures, such as the Sgr dwarf tidal stream and the Virgo and
Pisces overdensities in the halo, and the Monoceros stream closer to the
Galactic plane. Recent observational developments, based on accurate large-area
sky surveys, have made it abundantly clear that the Milky Way is a complex and
dynamic structure that is still being shaped by the infall (merging) of
neighboring smaller galaxies. Numerical simulations suggest that this merger
process plays a crucial role in establishing the structure and motions of stars
within galaxies, and is a generic feature of current cosmological models
\citep{SN2002,SL2003,Governato2004,Bullock2005,Brook2005,Governato2007,
Johnston2008,Font2011}. 

The main purpose of this review is to summarize some of the recent observational
progress in Milky Way studies, and the paradigm shifts\footnote{This phrase
was introduced by Thomas Kuhn, who apparently overused it (P. Yoachim, priv. comm.).
Here, it implies a change in the basic assumptions about galaxy formation,
from the smooth collapse model to the galaxy mergers scenario.} in our understanding of
galaxy formation and evolution resulting from this progress. This review is
focused on only a few studies, based mostly on data collected by the Sloan
Digital Sky Survey\footnote{www.sdss.org} described by \citet[][hereafter
SDSS]{SDSS}, and does not represent an exhaustive overview of all the progress
made during the last decade. One of our goals is to illustrate novel analysis
methods enabled by new datasets. We begin with a brief overview of methodology,
and of a few major datasets, and then describe the main observational results.
We conclude by discussing some of the unanswered questions, and observational
prospects for the immediate future.

\subsection{Stellar Populations: Definition and Role}

In astronomy, the term {stellar populations} is often associated with 
{Populations I, II, and III}, although the precise meanings of these populations
has changed over time. These stellar classes generally represent a
sequence of decreasing metallicity and increasing age. Here, we will use the
term ``stellar population'' to mean {any} collection of stars with common
spatial, kinematic, chemical, luminosity, and/or age distributions. For example,
a sample of red-giant stars selected using appropriate observables and selection
criteria is considered a population, although such a sample can include both
Population~I and Population~II stars. Similarly, we will often consider
populations of ``disk'' and ``halo'' stars, or samples selected from a narrow
color range. In summary, any sample of stars that share some common property
that is appropriate for mapping the Galaxy in the space of various observables
is hereafter considered to be a ``population''. 

Most studies of the Milky Way can be described as investigations of the stellar
distribution, or statistical behavior of various stellar populations, in the
seven-dimensional (7-D) phase space spanned by the three spatial coordinates,
three velocity components, and metallicity (of course, the abundances of
individual chemical elements can be treated as additional coordinates; this will
be a key ingredient for progress in the next decade). Depending on the quality,
quantity, and diversity of the data, such studies typically concentrate on only
a limited region of this 7-D space (e.g., the solar neighborhood, pencil beam
surveys, kinematically biased surveys), or consider only marginal distributions
(e.g., the number density of stars irrespective of their metallicity or
kinematics, proper-motion surveys without metallicity or radial-velocity
information). { The primary driver of the substantial progress in our
knowledge of the Milky Way over the last decade is the ability of modern sky
surveys to deliver the data required for determining the phase-space coordinates
for unprecedented numbers of faint stars over large areas of the sky.} For
example, in less than two decades the observational material for kinematic
mapping has progressed from the first pioneering studies based on only a few
hundred objects \citep{Majewski1992}, to over a thousand objects
\citep{ChibaBeers2000}, to the massive datasets including millions of stars
reviewed here.

The availability of large stellar samples enables detailed studies of various
distributions, including determination of the distributions' { shape}, rather
than considering only low-order statistical measures, as is done for small
samples. Deviations from Gaussian shapes often encode more information about the
history of galaxy assembly than the distribution's mean and dispersion. The
large samples are especially important for considering multi-variate
distributions (as opposed to one-dimensional marginal distributions), as the
so-called ``curse of dimensionality\footnote{When the problem dimensionality 
is high, the probability for a data point to belong to a multi-dimensional bin 
becomes small, and the bins become sparsely populated if the sample is 
not sufficiently large.}'' prevents their accurate determination with
small samples.  

In addition to increasing the sample size, the ability to detect faint stars is
crucial for extending the sample distance limit. With SDSS, it has become
possible to detect even main-sequence (dwarf) stars to a distance limit
exceeding 10 kpc, and thus to probe both the disk and halo populations {within
the same dataset}. The advantage of carrying out analyses of multiple
populations using stellar probes of similar intrinsic properties is difficult to
overstate. By way of comparison, the main-sequence stars in the Hipparcos sample
\citep{Perryman1997} only explore the volume within $\sim$100 pc of the Sun. The
primary advantage of main-sequence stars over probes such as RR Lyrae stars,
blue horizontal-branch (BHB) stars, and red-giant stars for studying Galactic
populations is that they are much more numerous (on the order of a thousand
times more than these other populations summed together), and thus enable a
substantially higher spatial resolution of the resulting phase-space maps
(assuming a fixed number of stars per multi-dimensional pixel in phase space,
and neglecting the intrinsic limit on spatial resolution set by the distance
precision). Of course, these other probes are still valuable, because they can
be used to explore the Galaxy to a larger distance limit than obtainable with
main-sequence stars alone. 

A theme common to most of the studies reviewed here is the use of photometric
parallax relations to estimate stellar distances, followed by the subsequent
{direct mapping} of various distributions using large samples of stars. This
mapping approach does not require {a-priori} model assumptions, and instead
{constructs multi-dimensional distribution maps first, and only then looks
for structure in the maps and compares them to Galactic models.} A key
observational breakthrough that made this approach possible was the availability
of accurate multi-band optical photometry to a faint flux limit over a large
area of sky, delivered by SDSS, as discussed below.

\subsection{Observations: Photometry, Spectroscopy, Astrometry} 

In order to determine the coordinates of a star in 7-D phase space, a variety of
astronomical techniques must be used. As always, the most crucial quantity to
measure is stellar distance. The largest sample of stars with trigonometric
distances, obtained by the Hipparcos survey, is too shallow (and too small) to
complement deep surveys such as SDSS and 2MASS (see below for an overview of
these surveys). Until the all-sky Gaia survey measures trigonometric distances
for about a billion stars brighter than $V=20$ (see \S 8.4),
various photometric methods need be employed in order to estimate distances. A
common aspect of these methods is that the luminosity (i.e., absolute magnitude)
of a star is determined by constraints derived from its color measurements, then
its distance is determined from the observed difference between its absolute and
apparent magnitude. For certain populations, for example RR Lyrae stars, a good
estimate of absolute magnitude is obtained as a simple constant (with some
metallicity dependence); for other populations, such as main-sequence stars, the
absolute magnitude depends on both effective temperature and metallicity, and
sometimes on age (or surface gravity) as well. A photometric parallax method for
main-sequence stars is described below. 

The most accurate measurements of stellar metallicity are based on spectroscopic
observations (but see below for a method of estimating metallicity using
photometric data alone). Spectroscopic measurements are especially important
when studying the extremely low end of the metallicity distribution function,
where photometric methods become insensitive. In addition to measuring chemical
composition, spectroscopic observations enable radial-velocity measurements. The
two largest existing stellar spectroscopic surveys are SDSS and RAVE (see the
next section). 

To measure all three components of the space-velocity vector, precise
astrometric observations are also required. The projection of the space-velocity
vector into the tangent plane (i.e., perpendicular to the radial-velocity
component) is measured using proper motion (the astrometric position shift per
unit time), which can be combined with the distance estimate to yield the space
velocity. The proper-motion measurements place an additional constraint on
observations; at least two (ideally, widely temporally separated) astrometric
epochs must be available. 

Therefore, multi-color imaging, multi-epoch astrometry, and spectroscopy are
required for measuring the coordinates of a star in the 7-D
position-velocity-metallicity phase space. It is the advent of massive and
accurate imaging and spectroscopic surveys that delivered such measurements for
large and relatively unbiased samples of stars, and thus enabled major progress
in Milky Way phase-space mapping during the last decade.

\subsubsection{A Photometric Parallax Method for Main-Sequence Stars \label{ppmethod}} 

In order to estimate distances to main-sequence stars with an accuracy of
10-20\% through the use photometric parallax relations, multi-band optical
photometry accurate to several percent (i.e., to several hundredths of a
magnitude) is required. This stringent requirement comes from the steepness of
the color-luminosity relation (the derivative of the absolute magnitude in the
SDSS $r-$band with respect to the $g-i$ color reaches $\sim$10 mag/mag at the
blue end), and is the main reason why it was not possible to use this method
with large sky surveys prior to SDSS. 

Using globular cluster data obtained in the SDSS photometric system, 
\citet[][hereafter Ivezic08]{Ivezic08} derived a polynomial expression for the absolute
magnitude of main-sequence stars in the $r$-band as a function of their $g-i$
color and metallicity (see their eqs. A2 and A7). The accuracy of the resulting
magnitudes is in the range 0.1-0.2 mag (Ivezic08; \citealt{SIJ08}), and the
method enables studies of the $\sim$100 pc to $\sim$10 kpc distance range when
used with SDSS data. { The ability to estimate distances to main-sequence stars
with sufficient accuracy using only SDSS photometry was crucial for wide-angle
panoramic mapping of the Galaxy to a distance limit 100 times farther than
possible with the Hipparcos data alone.}

\subsubsection{A Photometric Metallicity Method for Main-Sequence Stars \label{pmetmethod}} 

Stellar metallicity, together with effective temperature and surface gravity, is
one of the three main parameters that affect the observed spectral energy
distribution of most stars. In addition to being an informative observable for
deciphering the Milky Way's chemical history (e.g., \citealt{Majewski1993};  
\citealt{Freeman2002}; \citealt{Helmi08}; \citealt{Majewski2010}; and references therein),
knowledge of stellar metallicity is crucial for accurate estimates of distances using 
photometric parallax relations. 

The most accurate measurements of stellar metallicity are based on spectroscopic
observations. However, despite recent progress in the availability of at least
low-resolution digital stellar spectra (approaching a million!), the number of
stars detected in imaging surveys is still vastly larger. In addition to
generally providing better sky and depth coverage than spectroscopic surveys,
imaging surveys obtain essentially complete flux-limited samples of stars. These
simple selection criteria are advantageous when studying Galactic structure,
compared with the complex targeting criteria that are (by necessity) often used
for spectroscopic samples. 

As first suggested by \cite{SSH1955}, the depletion of metals in a stellar
atmosphere has a detectable effect on the emergent flux, in particular in the
blue spectral region where the density of metallic absorption lines is highest 
\citep[][and references therein]{BC2005}. Recent analysis of SDSS data by Ivezic08 demonstrated 
that for blue F- and G-type main-sequence stars, a metallicity estimate accurate
to $\sim$0.2 dex can be derived from the $u-g$ color. They derived a polynomial
expression that maps the measured $u-g$ vs. $g-r$ color space to effective
temperature and metallicity ($[Fe/H]$) \citep[for updated coefficients
see][hereafter Bond10]{Bond2010}. This transformation, applicable to stars with
$0.2 < g-r < 0.6$, was calibrated using $\sim$100,000 stars with available
spectroscopic metallicity determinations, and has errors in the range 0.2-0.3
dex when used with SDSS data (for stars in the range\footnote{This metallicity
range includes $\sim$99\% of all stars in the Milky Way.} $-2< [Fe/H] < +0.3$;
for more details see Ivezic08 and Bond10). Although applicable only within a
restricted color range, this calibration has enabled the construction of
metallicity maps using millions of stars, as discussed further below.

\section{THE ADVENT OF LARGE-AREA DIGITAL SKY SURVEYS}

Major advances in our understanding of the Milky Way have historically arisen
from dramatic improvements in our ability to ``see'', as vividly exemplified by
Galileo resolving the Milky Way disk into individual stars. Progressively larger
telescopes have been developed over the past century, but until recently most
astronomical investigations have focused on small samples of objects because the
largest telescope facilities typically have rather small fields of view, and
those with large fields of view could not detect very faint sources. Over the
past two decades, however, astronomy moved beyond the traditional observational
paradigm and undertook large-scale digital sky surveys, such as SDSS and the Two
Micron All Sky Survey\footnote{www.ipac.caltech.edu/2mass/} \citep[][hereafter
2MASS]{Skrutskie2006}. This observational progress, based on advances in
telescope construction, detectors, and above all, information technology, has
had a dramatic impact on nearly all fields of astronomy, including studies of
the structure of the Milky Way. Here we briefly overview the characteristics of
the most massive recent datasets.

\subsection{The SDSS Imaging and Spectroscopic Surveys \label{sdss}} 

The SDSS is a digital photometric and spectroscopic survey which covered over
one quarter of the Celestial Sphere in the North Galactic cap (approximately,
$b>30^\circ$), and produced a smaller area ($\sim$300 deg$^{2}$) but much deeper
survey in the Southern Galactic hemisphere, along the Celestial Equator
\citep[][and references therein]{Aihara2011}. The recent Data Release 8 has a
sky coverage of about 14,600 deg$^{2}$, and includes photometric measurements
for 469 million unique objects (approximately half are stars). The completeness
of the SDSS catalogs for point sources is $\sim$99\% at the bright end, dropping
to 95\% at an $r$-band magnitude of $\sim$22. The wavelength coverage of the
SDSS photometric system ($ugriz$, with effective wavelengths from 3540 \AA\ to
9250 \AA), and photometry accurate to $\sim$0.02 mag, have enabled photometric
parallax and metallicity estimates for many millions of stars. For comparison,
the best large-area optical sky survey prior to SDSS, the photographic Palomar
Observatory Sky Survey, had only two photometric bands, several times larger
photometric errors, and was limited by uncertain zero points that varied from
plate-to-plate \citep{Sesar2006}. 

\begin{figure}%
\centerline{\psfig{figure=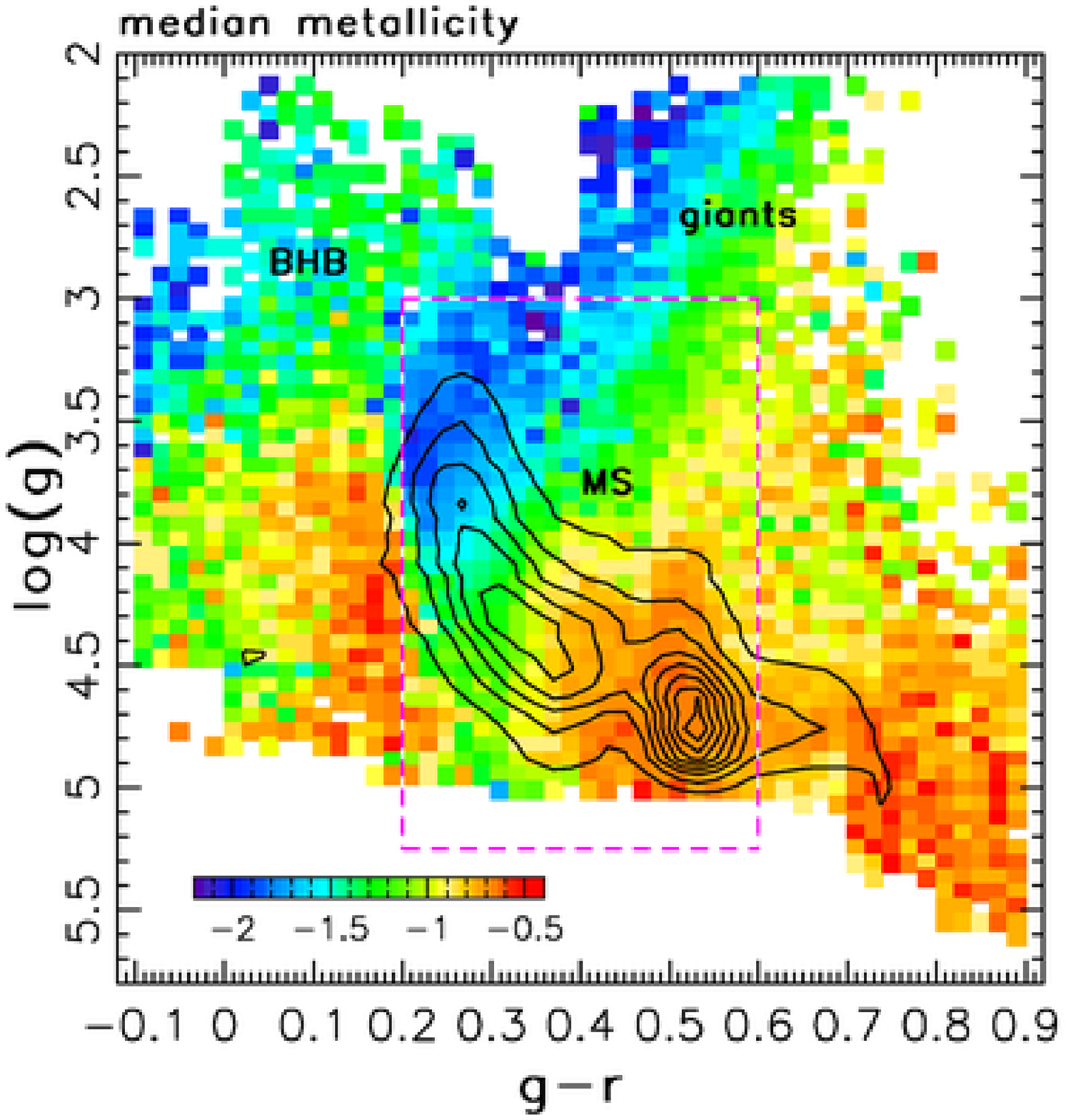,height=25pc}}
\vskip -0.2in 
\caption{The stellar content of the SDSS spectroscopic surveys, through Data Release
6 (Figure 1 from \citealt{Ivezic08}). Linearly spaced contours show  the distribution of
$\sim$110,000 stars with $g < 19.5$ and $0.1<g-r<0.9$ (corresponding to
effective temperatures in the range 4500~K $< T_{\rm eff} < $ 8200~K) in the
$log(g)$ vs. $g-r$ plane ($g$ is the SDSS $g-$band magnitude, and $log(g)$ measures the
surface gravity). The multimodal distribution is a result of the SDSS
target selection algorithm. The color scheme shows the median metallicity for all
0.02 mag by 0.06 dex pixels that contain at least 10 stars. The fraction
of stars with $log(g) < 3$ (giants) is 4\%, and they are mostly found in two
color regions: $-0.1<g-r<0.2$ (blue horizontal branch, BHB, stars) and $0.4<g-r<0.65$
(red giants). They are dominated by low-metallicity stars ($[Fe/H] < -1$). The dashed lines 
outline the main-sequence (MS) region, where photometric metallicity methods can be
applied.}
\label{TIIf1}
\end{figure}

In addition to its imaging survey data, SDSS has obtained well over half a
million stellar spectra, many as part of the Sloan Extension for Galactic
Understanding and Exploration \citep[SEGUE;][]{Yanny09}, and its continuation
SEGUE-2, sub-surveys carried out during the first (SDSS-II) and second
\citep[SDSS-III;][]{SDSSIII} extensions of the
SDSS project. These spectra have wavelength coverage 3800--9200~\AA~and spectral
resolving power $R \sim$$2000$, with a typical signal-to-noise ratio per 150 km
s$^{-1}$ resolution element of $>30$ at $r\sim18.5$, and $\sim$3 at $r\sim20$.
SDSS stellar spectra are of sufficient quality to provide robust and accurate
stellar parameters, such as effective temperature, surface gravity, and
metallicity (parameterized as $[Fe/H]$). These publicly available parameters are
estimated using a variety of methods implemented in an automated pipeline
\citep[][the SEGUE Stellar Parameters Pipeline, SSPP]{Beers2006}. A detailed
discussion of these methods and their performance can be found in Papers I-V of
the SSPP series \citep{Lee2008a,Lee2008b,APOGEE2008,Smolinski2011,Lee2011segue}. 
Based on a comparison with high-resolution abundance determinations,
they demonstrate that the combination of spectroscopy and photometry from SDSS
is capable of delivering estimates of $T_{\rm eff}$, $log(g)$, and $[Fe/H]$
accurate to 200~K (3\%), 0.3 dex, and 0.2 dex, respectively. Random errors for
the radial-velocity measurements are a function of spectral type, but are
usually $< 5$~{\kms} for stars brighter than $r\sim18$, rising to
$\sim$$20$~{\kms} for stars with $r\sim20$ \citep{Pourbaix2005,Yanny09}.
\cite{Lee2011segue} demonstrate that SDSS spectra are of sufficient quality to
also determine $[\alpha/Fe]$ with errors below 0.1 dex (for stars with
temperatures in the range 4500-7000 K and sufficient signal-to-noise ratios).
The distribution of SDSS stars with available spectroscopic estimates of
atmospheric parameters in the log(g) vs. color plane is shown in
Figure~\ref{TIIf1}.

\subsection{The SDSS-POSS Proper-Motion Survey \label{sdssposs}}

The time difference of about half a century between the Palomar Observatory Sky
Survey (POSS) and the SDSS imaging observations provides an excellent baseline
to measure proper motions for tens of millions of stars to faint brightness
levels. \cite{Munn04} addressed the problem of large systematic astrometric
errors in the POSS catalogs by recalibrating the USNO-B catalog
\citep{Monet2003}, using the positions of galaxies measured by SDSS. As a result
of this calibration, the SDSS-POSS proper-motion measurements are now available
for about 100 million unresolved sources, most of them stars. This catalog also
includes about 70,000 spectroscopically-confirmed SDSS quasars that were used to
robustly estimate the proper-motion errors \citep{Bond2010}. The random errors
increase from $\sim$$3$~mas~yr$^{-1}$ at the bright end to
$\sim$$6$~mas~yr$^{-1}$ at $r\sim20$ (the sample completeness limit), with
systematic errors that are typically an order of magnitude smaller, and with
very small variation across the sky (for a discussion of deviations from
Gaussian error behavior, see \citealt{Dong2011}). Even for stars at distances of
1 kpc, the implied tangential velocity errors are as small as 10-20 km s$^{-1}$,
and well matched to the SDSS radial velocity accuracy. This catalog represents a
major improvement over previously available data sets {both in size and
accuracy}.

\subsection{The 2MASS Imaging Survey}

The 2MASS databases are derived from an all-sky near-IR photometric survey with
limiting (Vega-based, 10$\sigma$) magnitudes of $J$=15.8, $H$=15.1, and
$K$=14.3. The 2MASS point source catalog contains positional and photometric
information for 471 million sources (mostly stars). The near-IR 2MASS colors are
not as good as the optical SDSS colors for estimating photometric parallax and
metallicity, because they only probe the Rayleigh-Jeans tail of the stellar
spectral energy distribution. On the other hand, a major advantage of 2MASS over
SDSS is the full sky coverage, and its ability to penetrate deeper through the
interstellar dust in the Galactic plane. In addition, it is much easier to
photometrically identify certain stellar populations using near-IR data than
with optical data. For example, \cite{Majewski2003} have demonstrated that
M-giant candidates color selected from the 2MASS database are extremely powerful
probes of halo substructure out to $\sim$100 kpc over the entire sky (these
stars are practically impossible to robustly identify using SDSS photometry).
For an analysis of the joint SDSS-2MASS stellar dataset, we refer the reader to
\cite{Covey2007}.

\subsection{The RAVE Spectroscopic Survey}

The Radial Velocity Experiment\footnote{www.rave-survey.aip.de} (RAVE) is a
major new spectroscopic survey aiming to measure radial velocities and stellar
atmosphere parameters (temperature, surface gravity, and metallicity) of up to
one million stars using the Six Degree Field multi-object spectrograph on the
1.2m UK Schmidt Telescope of the Anglo-Australian Observatory
\citep{Steinmetz2006}. RAVE stars are selected from the magnitude range $9 < I <
12$, and represent a bright complement to the SDSS spectroscopic sample
\citep{Siebert2011}. The wavelength range for the RAVE spectra (8410--8795~\AA,
in the region of the Ca II Triplet, with a spectral resolving power of $R
\sim8000$) includes a number of lines in addition to calcium, and should
eventually provide reliable estimates of $[\alpha/Fe]$ for numerous stars, in
addition to overall metallicity ($[Fe/H]$). The RAVE catalog of stellar
elemental abundances \citep{Boeche2011} includes estimates of abundances for Mg,
Al, Si, Ca, Ti, Fe and Ni, with a mean error of $\sim$0.2 dex, for some 36,000
stars.

The third RAVE data release includes radial-velocity data for $\sim$77,000 stars
and stellar parameters for $\sim$40,000 stars \citep{Siebert2011}, but spectra
are already collected for over 300,000 stars \citep{Zwitter2010}. With a 
radial-velocity error of about 2 km $s^{-1}$, the RAVE velocities are more accurate
than those delivered by SDSS, and are well-suited for detailed kinematic studies
of nearby disk stars \citep{Ruchti2011}. Proper motions (of varying accuracy)
are available for most of the RAVE stars from other surveys, and model-based
distance determinations accurate to $\sim$20\% are also available
\citep{Zwitter2010,Burnett2011}. 

\begin{figure}%
\vskip -1.9in 
\centerline{\psfig{figure=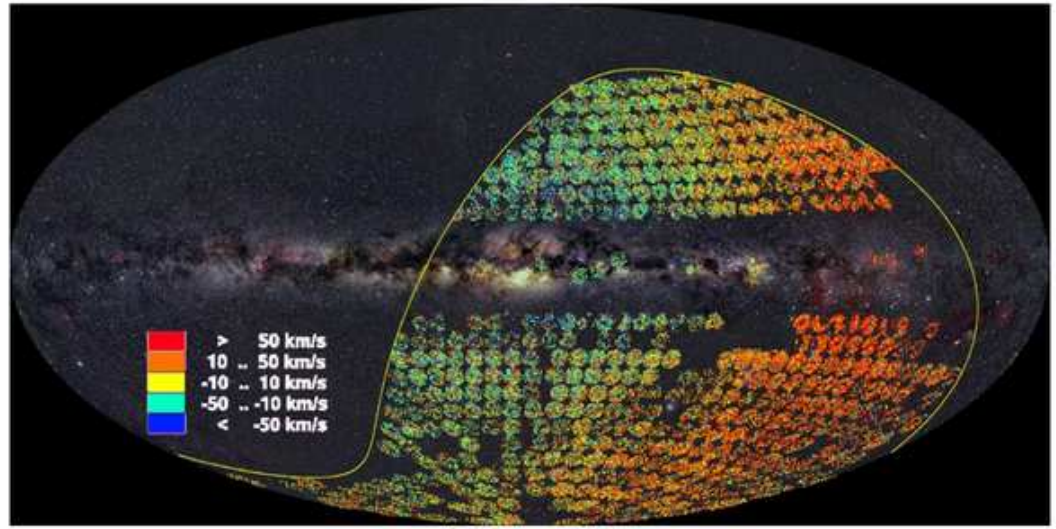,height=45pc}}
\vskip -1.9in 
\caption{The sky coverage of RAVE DR3, shown as an Aitoff
projection in Galactic coordinates and color-coded by the mean radial 
velocity \citep[Figure 17 from][]{Siebert2011}. The nearly contiguous 
coverage over a wide solid angle, with detailed data for (eventually) up
to a million stars is representative of modern Milky Way surveys and
enables new approaches to studying the Galaxy. The distances probed 
by RAVE stars ranges up to $\sim$1 kpc, thus the RAVE dataset provides
a valuable link between the nearby Hipparcos sample ($<100$ pc) and the 
more distant SDSS sample.}
\label{RAVE}
\end{figure}

The distances probed by RAVE stars range from $\sim$300 pc (dwarfs) to
$\sim$1-2 kpc (giants), and thus the RAVE dataset ``connects'' the nearby
Hipparcos sample and the more distant SDSS sample. Due to these distance limits,
RAVE data are more relevant for disk than for halo investigations. However, the
RAVE survey has demonstrated the ability to identify at least a limited number
(hundreds in the present sample, eventually several thousand) of bright, very
low-metallicity stars with $[Fe/H] < -2.0$, including a handful with metallicity
as low as $[Fe/H] = -4.0$ \citep{Fulbright2010}. Bright very metal-poor stars
are of particular interest, since follow-up high-resolution spectroscopy can be
obtained with relatively short integration times on 4m to 8m-class telescopes.
In addition, the large-area nearly-contiguous coverage of RAVE survey (see
Figure~\ref{RAVE}) is very useful for panoramic Galactic mapping.

\section{OVERVIEW OF THE STATE-OF-THE-ART A DECADE AGO \label{decadeAgo}}

Before discussing results concerning the nature of Galactic structure and
stellar populations obtained during the last decade, we briefly review
the state of related knowledge a decade ago.  We concentrate on the
spectroscopic surveys in existence at that time, and the questions they sought
to address.

Circa 2000, there were two primary approaches in common use for the detection
and analysis of significant numbers of stars with membership in the thick-disk and
halo populations of the Galaxy. The first, essentially a continuation of the
high-proper-motion based surveys pioneered by Sandage and colleagues, is
exemplified by the work of \cite{RyanNorris1991}
and that of 
\cite{Carney1996}, and references therein. Both of these works
concentrated on members of the halo population, although stars from the disk
system were certainly present in their samples as well. The second, which
followed on the efforts of \cite{Norris1986}, was the assemblage \citep{Beers2000}
and analysis \citep{ChibaBeers2000} of a large sample of non-kinematically
selected stars with [Fe/H] $< -0.6$ (this metallicity limit was chosen
to minimize the contribution of disk-system stars).  It was considered of central importance
(as it remains now) to contrast the derived properties of samples chosen with
differing biases -- the former on kinematics, and the latter on metallicity.
For this reason, papers that followed commonly adopted the halo metallicity
distribution functions (MDFs) as derived
from the kinematically selected samples, and the kinematics of the thick-disk
and halo populations based on the non-kinematically selected samples. It is worth
noting that, even prior to the results obtained by these two surveys, \cite{Freeman1987}
called attention to the puzzling differences in the trends of the derived
halo rotation velocities and velocity dispersions as a function of declining
metallicity for the extant kinematically vs. non-kinematically selected stars. 
Today, it seems likely that this puzzle may be resolved by the recognition that,
even within the halo component, their exists a strong coupling between kinematics
and metallicity that was not previously apparent.

The most pressing questions from a decade ago included the following, asked and
answered (even if only partially) below. Answers to the above questions were
essentially all limited by the relatively small numbers of stars with 7-D
phase-space information then available, as well as by the troublesome selection
biases that were known to exist in the tracer samples. Even so, some progress
toward resolution of these issues was being made at the time. 

\subsection{What is the nature of the halo MDF?}

What is nature of the MDF for the halo population, and does it include significant 
numbers of stars with metallicities below [Fe/H] $\sim -3.0$?

Although the proper-motion-selected samples included only a handful of stars
with $[Fe/H] < -3.0$, the objective-prism surveys that were the source of the
lowest metallicity stars known at that time \citep[e.g.,][]{Beers1992} included
tens of stars below this metallicity. The inference could be made that, although
stars of such low metallicity were rare compared to the more metal-rich halo
stars with $[Fe/H] \sim -1.5$, they did in fact exist in substantial numbers. 

\subsection{The ELS vs. SZ model of halo formation}

Can the kinematics of the halo of the Galaxy be adequately described by the rapid 
collapse model of \cite{ELS62}, ELS, or did the observations require the more extended, 
chaotic assembly picture of \cite{SearleZinn}?

One of the primary discriminants between the suggested galaxy formation models
is the presence (or not) of halo stars with low orbital eccentricities, which
might not be expected to be found in significant number if the ELS model was the
correct interpretation. Papers published prior to \cite{ChibaBeers2000} did in
fact identify such stars \citep[e.g.,][]{NBP1985}, but only relatively few. The
question was essentially resolved by the substantially larger numbers of
low-metallicity, low-eccentricity halo stars discussed by \cite{ChibaBeers2000}.

\subsection{What is the shape of the halo?} 

What is the shape of the density profile of the stellar halo,
and does it remain constant with increasing distance (and/or declining
metallicity)?

A number of papers prior to 2000 \citep[e.g.,][]{Hartwick1987,Preston1991}
pointed out that the shape of the stellar halo density profile changes with
Galactocentric distance, in the sense that it is relatively flattened in the
inner region, and becomes substantially rounder in the outer region. This result
was validated by inferences based on the kinematics of local halo stars carried
out by \cite{SLZ1990} and \cite{ChibaBeers2000}. 

\subsection{Are the thin and thick disk distinct entities?}

Is the disk system adequately modeled by the superposition of
a thin-disk and thick-disk population, that is, are these two components
demonstrably distinct from one another?

Although \cite{NorrisRyan1991} made the argument that a continuous extended disk
configuration could be supported by the existing data at that time,
\cite{ChibaBeers2000} claimed that the kinematics and abundances of thick-disk
stars indicated that a distinct thick-disk component was the more likely
interpretation. However, the question remained basically open as of a decade
ago. 
 
\subsection{Is the metal-weak thick disk real?}

Is there evidence for the additional presence of a metal-weak thick-disk (MWTD) 
population, rotationally supported, but
extending to lower metallicity stars than the canonical thick disk?
 
Although the original suggestion that a MWTD thick-disk component may exist
\citep{NBP1985,Morrison1990} was supported by the analysis of
\cite{ChibaBeers2000}, claims to the contrary based on revised photometric
studies \citep{Twarog1994} and high-resolution spectroscopic studies of
suggested MWTD stars \citep[e.g.,][]{RyanLambert1995} (but see also
\citealt{Bonifacio1999}) called the reality of the MWTD into question, at least
at that time. 

\subsection{Are there stellar streams other than Sagittarius?} 

Can one identify other streams of stars, similar to the then-recognized Sgr stream,
that might be associated with origin in stripped dwarf galaxies?

\cite{Helmi1999} presented the detection of a small number of
halo stars that appeared as a statistically significant overdensity in an
otherwise sparsely population region of angular momentum phase space, and argued
that they may have originated from the stripping of a once-coherent structure
such as a dwarf galaxy. This inference was supported by additional data from
\cite{ChibaBeers2000}, and others since, but of  course it was only one such
example.

So, in toto, among the issues mentioned above, the data available as of a decade
ago could at best claim only one clear ``victory" (Issue 2), three strong
``maybes" (Issues 1, 3, and 6), and two ``yet to be decided" (Issues 4 and 5).
Larger samples with well-understood selection biases were clearly needed.

\section{WHAT DID WE LEARN DURING THE LAST DECADE?}

Until recently, our global view of the Milky Way was hampered by the fact that
most detected stars had no reliable distance estimates. Those stars that had
usable distances were either limited to the solar neighborhood (e.g., for
main-sequence stars in the Hipparcos sample to within $\sim$100 pc, or only
$\sim$1\% of our distance to the Galactic center), or to smaller pencil-beam
surveys. Our knowledge of the basic structural components of the Milky Way was
thus limited to indirect inferences based on stellar population models motivated
by other spiral galaxies \citep[e.g.,][]{BS1980,Robin03}. This limitation was
alleviated recently by the advent of SDSS, which provided accurate digital
multi-band optical photometry across a quarter of the sky. The SDSS photometry
enabled the development and application of photometric parallax methods, which in
turn led to direct mapping of stellar distributions in the multi-dimensional
space spanned by spatial coordinates, velocity components, and chemical
abundance measurements. The resulting maps provided the quantitative basis for
separating the main structural components of the Galaxy, obtaining information
on their phenomenological description, and enabled efficient searches for
substructure and a robust comparison with various model predictions. 

We first describe how these new data clearly reveal the disk and halo as two
distinct Galaxy components, and then describe each of them in more detail. We do
not discuss in detail here the third major Galaxy component, the bulge. For a
recent excellent review of the bulge, see \cite{MZ2008bulge}. It is expected
that spectroscopic data being collected by the SDSS-III Apache Point Observatory
Galaxy Evolution Experiment (APOGEE) will soon yield unprecedented insight into
chemical and kinematic properties of the bulge \citep[][see also
\S\ref{Sec:APOGEE}]{SDSSIII}. We do not discuss the properties of the Galactic
bar here, and refer the reader to recent studies by \citet{CL2008},
\citet{Rat2007}, and \citet{Robin2011}, which summarize 
the current state of the art and include relevant references.

\subsection{Separation of the Main Structural Components}

\begin{figure}%
\centerline{\psfig{figure=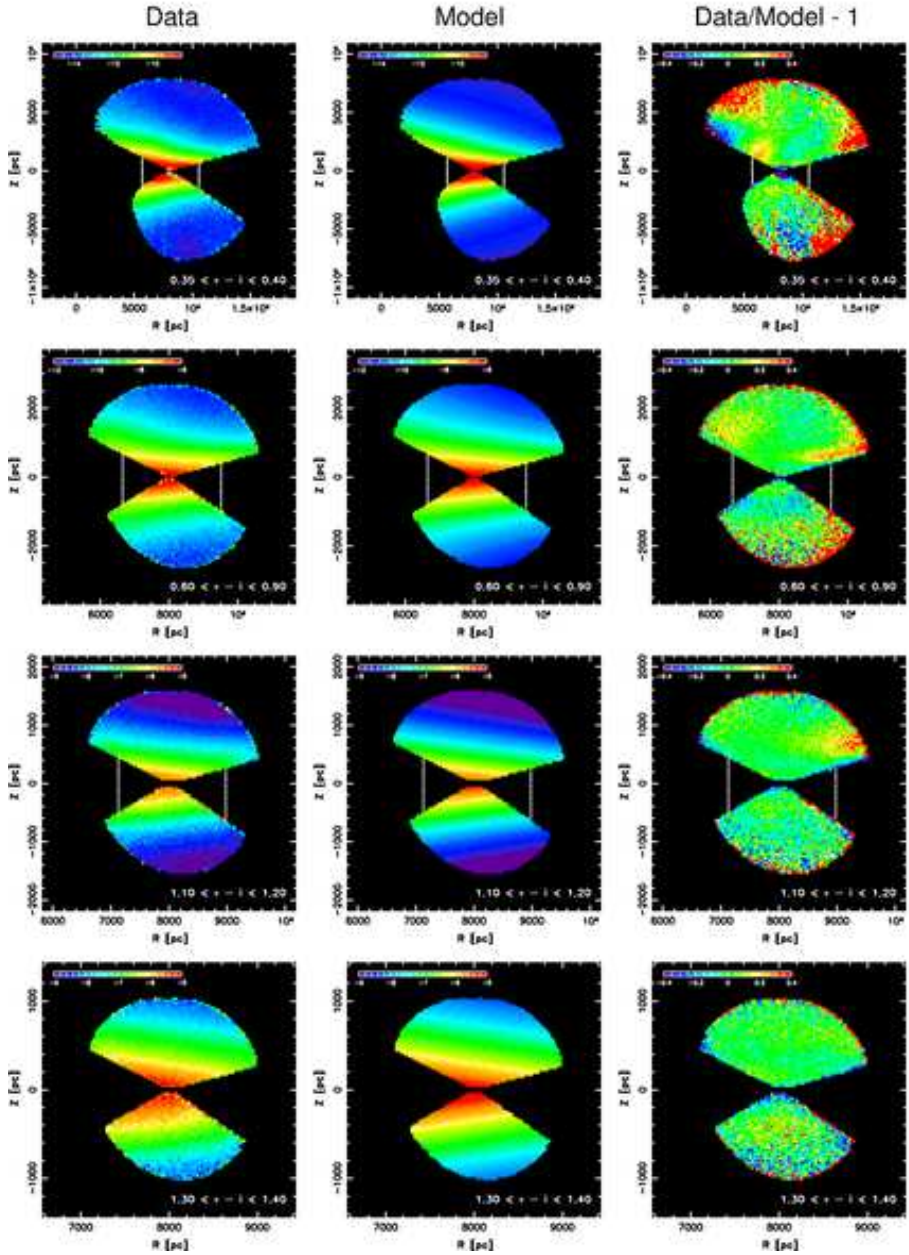,height=40pc}}
\vskip -0.3in 
\caption{Figure 26 from \cite{Juric08}. The panels in the left column
show the measured stellar number density, as a function of Galactic cylindrical
coordinates, for stars selected from narrow ranges of $r-i$ color ($0.35< r-i
<0.40$ in the top row to $1.30< r-i<1.40$ in the bottom row). The panels in the
middle column show the best-fit smooth models; panels in the right column show
the normalized (data-model) difference map. Note the large overdensities visible
in the top three panels in the right column.}
\label{TIf26}
\end{figure}

Before the disk and halo can be studied in detail, a robust and accurate
scheme for classifying stars into these two components needs to be
developed. Using photometric data for $\sim$50 million stars,
\citet[][hereafter Juric08]{Juric08} constructed 3-dimensional maps (data
cubes) of the stellar number-density distribution for 19 narrow color bins
that span spectral types from mid-F to early M-type stars. When the bin
color is varied from the reddest to the bluest one, the maps are ``zoomed
out'', with subsamples covering distances ranging from 100 pc to 15 kpc.
The distance to each star was estimated using a maximum likelihood
implementation of the photometric parallax method, and stars are binned and
counted in small 3-dimensional pixels whose size depends on the dynamical
range provided by each color bin and Poisson noise limits (typically there
are 250,000 pixels per map). Examples of two-dimensional projections of the
resulting maps are shown in Figure~\ref{TIf26}. 

These maps are a powerful tool for studying the Milky Way's stellar number
density distribution. Traditional methods for modeling stellar counts in
the magnitude-color space need to adopt a large number of poorly-known
relations, such as the stellar initial mass function, stellar
mass-luminosity relationship, stellar luminosity function, and a geometric
description of the postulated components such as the disks, bulge, and halo.
Alternatively, with these number-density maps the Milky Way's structure can
be examined without any a-priori assumptions about its components: {The
analysis of the Milky Way's structure is then akin to studies of external
galaxies.}

The quantitative description of these maps is still a non-trivial task, due
to the presence of rich substructures within the components. While halo
substructure has been known for some time \citep{ive00,yan00,Majewski2003,
vz06, Belo2006a}, these new maps demonstrate that disk substructure is also
complex. Nevertheless, the gross behavior can be captured by assuming
standard Galaxy models based on two exponential disks and a power-law halo.
Juric08 determined the best-fit parameter values for full two-dimensional
smooth models, and further refined them using residual minimization
algorithms (see Table~\ref{Tab:Juric08model}).

\begin{table}%
\def~{\hphantom{0}}
\caption{The Best-fit Parameters for the Juric08 Galactic Model$^a$}
\vskip 0.2in
\begin{tabular}{@{}lrrr@{}}%
\toprule
Parameter & Measured & Bias-corrected & Error estimate  \\ 
\colrule
$Z_0$  &  25		&  \dots	        & $20$\%	\\
$L_1$  &  2150	& 2600		& $20$\%	\\
$H_1$  &  245		& 300		& $20$\%	\\
$f_d$   &  0.13		& 0.12		& $10$\%	\\
$L_2$  &  3261	& 3600		& $20$\%	\\
$H_2$ &  743		& 900		& $20$\%	\\
$f_h$   &  0.0051	&  \dots	        & $25$\%	\\
$q$     &  0.64		&  \dots    	& $\le 0.1$ 	\\
$n$     &  2.77		&  \dots  	& $\le 0.2$      \\ 	
\botrule
\end{tabular}
\label{Tab:Juric08model}
\\
\vskip 0.1in
{$^a$ Best-fit Galactic model parameters from Juric08, as directly measured
from the apparent number-density distribution maps (2$^{\rm nd}$ column),
after correcting for a 35\% assumed binary fraction and Malmquist bias
due to photometric errors and dispersion around the mean of the photometric
paralax relation (3$^{\rm rd}$ column). $Z_\odot$ is the Solar offset from
the Galactic plane (pc), $H_1$, $H_2$ and $L_1$, $L_2$ (pc) are the
scale heights and lengths for the thin and thick disk, respectively, $f_d$
and $f_h$ are the thick disk and halo normalizations relative to the thin
disk at ($R=R_\odot, Z=0$), $q$ parametrizes the halo ellipticity (with the
ellipsoid described by axes $a=b$ and $c=q\,a$; for $q<1$ the halo is
oblate, that is, ``squashed'' in the same sense as the disk), and $n$ is
the power-law index for the halo number-density profile.}
\end{table}

\begin{figure}%
\vskip -0.3in 
\centerline{\psfig{figure=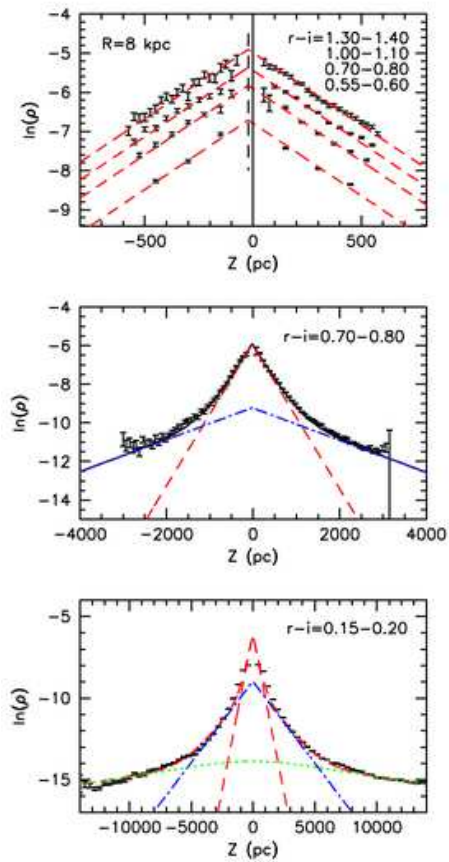,height=30pc}}
\vskip -0.6in 
\caption{Cross sections through maps similar to those shown in
Figure~\ref{TIf26}, showing the vertical ($|Z|$) distribution at $R=8$ kpc and
for different $r-i$ color bins \citep[Figure 15 from][]{Juric08}. The lines are
exponential models fitted to the points (the { sech$^2$} function is not a good
fit; see footnote 28 in Juric08). The dashed lines in the top panel
correspond to a fit with a single, exponential disk. The dashed line in the
middle panel corresponds to a sum of two disks with scale heights of 270 pc and
1200 pc, respectively, and a relative normalization of 0.04 (the ``thin'' and
the ``thick'' disks). The dashed line in the bottom panel (closely following the
data points) corresponds to a sum of two disks and a power-law spherical halo.
The dashed line and the dot-dashed line are the disk contributions, and the halo
contribution is shown by the dotted line. For the best-fit
parameters see Table~\ref{Tab:Juric08model}.} 
\label{TIf15}
\end{figure}

A cross section of the maps from Figure~\ref{TIf26} in the direction
perpendicular to the disk plane is shown in Figure~\ref{TIf15}. The data
shown in the middle and bottom panels clearly confirm a change in the
number-count behavior around $|Z|\sim$1-1.5 kpc, interpreted as evidence
for an extended ``thick'' disk component by \cite{Yoshii1982} (who referred
to it as a ``halo'' component, even though its inferred density was ten
times that of the local halo and its scale height of $\sim 2$ kpc was
commensurate with the values later determined by others) and
\cite{GilmoreReid1983}. At the point where the additional, more-extended
component becomes unable to explain the star counts, around $|Z|\sim$5 kpc,
another component -- the stellar halo -- is invoked to explain the data.
Although these modern counts have exceedingly low statistical noise and
fairly well-understood systematics, the three-component fit to the data
shown in the bottom panel begs the question whether a single-component fit
with some other function, parametrized with fewer free parameters, might
suffice.

\begin{figure}%
\vskip -1.0in
\centerline{\psfig{figure=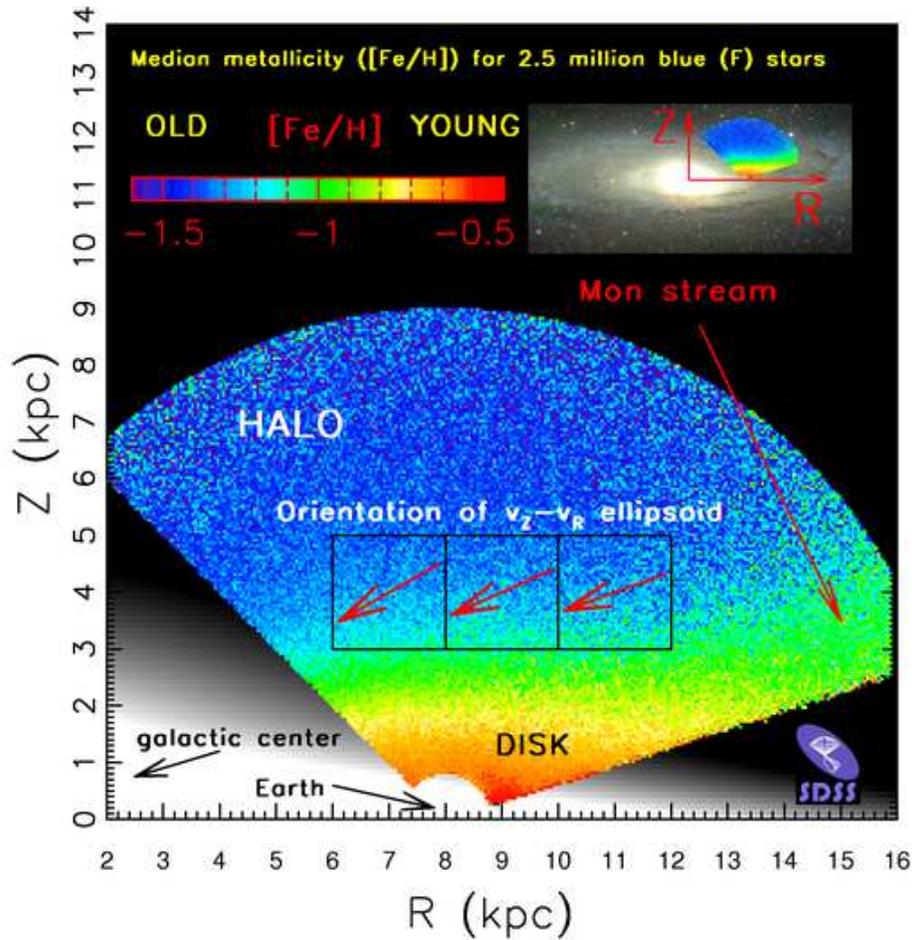,height=45pc}}
\vskip -2.2in 
\caption{
Variation of the median photometric stellar metallicity for $\sim$2.5 million
stars from SDSS with $14.5 < r <20$ and $0.2 < g-r < 0.4$, and photometric
distance in the 0.8-9 kpc range, in cylindrical Galactic coordinates $R$ and
$|Z|$ (adapted from Ivezic08). The $\sim$40,000 pixels (50 pc by 50 pc) contained 
in this map are colored according to the legend in the top left. Note that the gradient 
of the median metallicity is essentially parallel to the $|Z|$ axis, except in the
region of the Monoceros stream, as marked. The gray-scale background is the
best-fit model for the stellar number-density distribution from Juric08. The inset
in the top right illustrates the extent of the data volume relative to the rest
of Galaxy; the background image is the Andromeda galaxy. The three squares 
outline the regions used to construct the $v_Z-v_R$ ellipsoids shown in 
Figure~\ref{TIIIf13}. The arrows illustrate the variation of the ellipsoid orientation, 
which always points towards the Galactic center.}
\label{FeHmap}
\end{figure}

It turns out that the three additive components invoked to explain the
counts exhibit distinctive chemical and kinematic behavior as well.
Figure~\ref{FeHmap} shows a panoramic view of the variation in the median
$[Fe/H]$ over an unprecedentedly large volume of the Galaxy. The map is
based on photometric metallicity estimates for a sample of 2.5 million blue
main-sequence stars (most of F spectral type) selected using very simple
color and flux limits. It is easily discernible that the median metallicity
farther than $\sim5$ kpc from the Galactic plane is very uniform and about
1 dex lower than for stars within $\sim1$ kpc from the plane. 
 
\begin{figure}%
\vskip -1.9in
\centerline{\psfig{figure=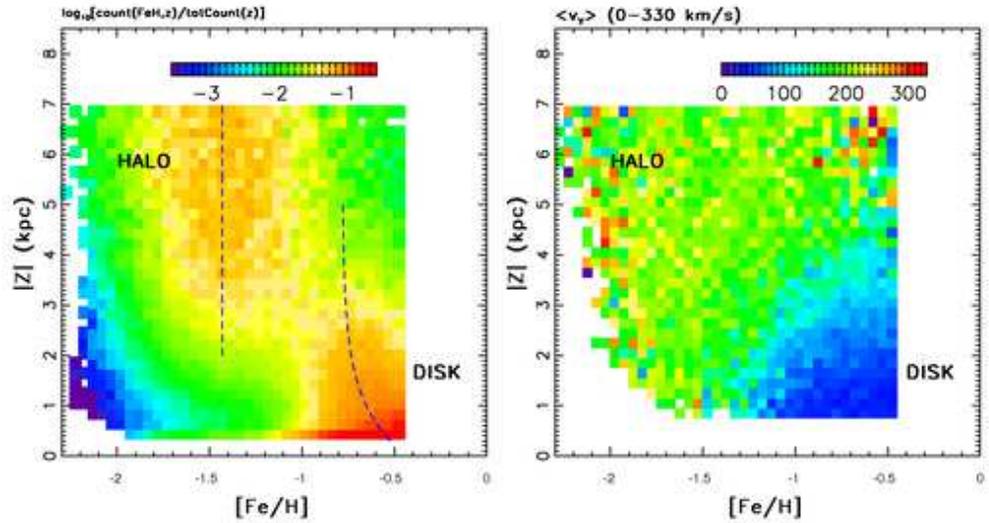,height=49pc}}
\vskip -1.6in 
\caption{Figure 9 from \cite{Ivezic08}. 
The left panel shows the conditional metallicity probability distribution (each
row of pixels integrates to unity) for $\sim$60,000 stars from a cylinder
perpendicular to the Galactic plane, centered on the Sun, and with a radius of 1
kpc. The values are color coded on a logarithmic scale according to the legend
on top. The lack of stars with $[Fe/H]>-0.5$ is due to a bias in SDSS Data Release
6 reductions, and an updated version of this map based on Data Release 7 is 
shown in Figure A.3 from \cite{Bond2010}. The right panel shows the median heliocentric rotational
velocity component (the value of $\sim$220 km s$^{-1}$ corresponds to no
rotation), as a function of metallicity and distance from the Galactic
plane, for the $\sim$40,000 stars from the left panel that also satisfy $|b| > 80^\circ$.}
\label{TIIf9}
\end{figure}

The reason for the rapid decrease of median stellar metallicity with $|Z|$
for $|Z|<5$ kpc, and very little variation farther from the plane, is
illustrated in the left panel in Figure~\ref{TIIf9}. The two distinct
distributions imply different Galaxy components, the halo and the disk, and
are clearly evident. High-metallicity disk stars dominate close to the
plane, while low-metallicity halo stars dominate beyond 3 kpc from the
plane. The median metallicity for disk stars exhibit a vertical gradient,
while the halo stars (at least in the relatively nearby volume) have a
spatially invariant metallicity distribution. As $|Z|$ increases from
$|Z|\sim$2 kpc to $|Z|\sim$4 kpc, halo stars become more numerous than disk
stars, and the median metallicity drops by $\sim$1.0 to 1.5 dex. A more detailed
and quantitative discussion of these metallicity distributions can be found
in Ivezic08. 

\begin{figure}%
\vskip -1.5in
\centerline{\psfig{figure=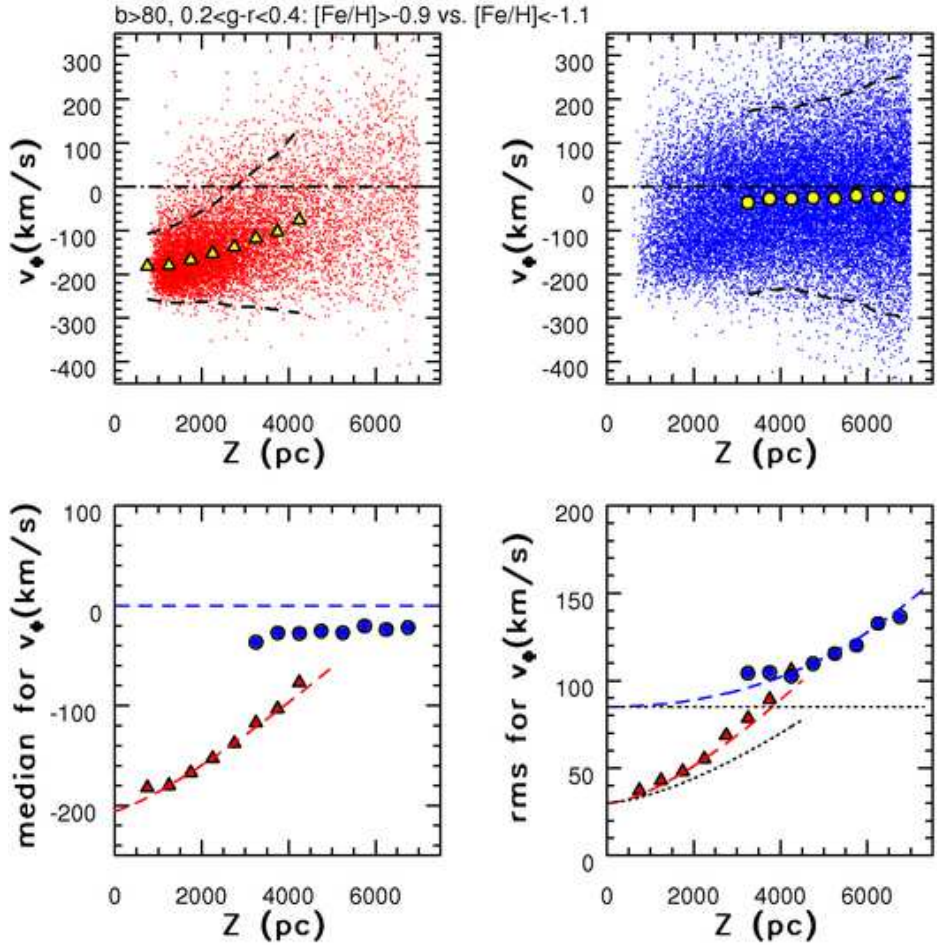,height=43pc}}
\vskip -0.6in 
\caption{Figure 5 from \cite{Bond2010}. A comparison of the variation of
rotational velocity (see their eqn.~8), $v_\Phi$, on distance from the Galactic plane, $|Z|$,
for 14,000 high-metallicity ($[Fe/H] > -0.9$; top left panel) and 23,000
low-metallicity ($[Fe/H] < -1.1$; top right panel) stars with $|b| > 80^\circ$. In the top
two panels, individual stars are plotted as small dots, and the medians in bins
of $|Z|$ are plotted as large symbols. The $2\sigma$ envelope around the medians
is shown by dashed lines. The bottom two panels compare the medians (left) and
dispersions (right) for the two subsamples shown in the top panels, and the
dashed lines in the bottom two panels show predictions of a kinematic model. The
dotted lines in the bottom-right panel show model dispersions (without
correction for measurement errors).}
\label{TIIIf5}
\end{figure}

These two components, with distinct metallicity distributions, also have
vastly different kinematic behavior, as shown in the right panel in
Figure~\ref{TIIf9}. The high-metallicity disk stars exhibit large
rotational velocity (about 220 km s$^{-1}$), while the low-metallicity halo
stars display behavior consistent with no net rotation (to within 10-20 km
s$^{-1}$). Similar to the behavior of their metallicity distributions, the
rotational velocity for disk stars decreases with the distance from the
Galactic plane, while it remains constant for nearby halo stars (see
Figure~\ref{TIIIf5}). 

{Therefore, reasonably clean subsamples of halo and disk stars can be
defined using a simple metallicity boundary $[Fe/H] = -1$.} We proceed
below with a more detailed discussions of each component.

\section{THE MILKY WAY DISK} 

Recent massive datasets based on SDSS have confirmed, with exceedingly high
statistical signal-to-noise ratios, the abrupt change of slope in the log(counts)
vs. $|Z|$ plot around $|Z|\sim1$ kpc for disk stars (Juric08). This slope
change was discovered almost three decades ago, and interpreted as evidence
for two disk components: the thin disk and the thick disk. Over a similar
range in $|Z|$, there are clearly detected vertical gradients in the median
stellar metallicity and rotational velocity (Ivezic08, and references
therein). A key question now is whether the two disk components required to
explain the counts can also be used to account for the chemical and
kinematic measurements for the same stars. For example, are the metallicity
and kinematic gradients due to the interplay of two {additive} components
(with thick-disk stars dominating beyond $|Z|\sim1$), or do they instead
reveal a single disk with { complex} variations of basic properties
(perhaps driven by a hidden variable, such as age)? In other words, do the
new data require a disk decomposition into thin- and thick-disk components,
and if so, what is an optimal way to define these components? It turns out
that, even with the new data collected over the last decade, it is not easy
to answer these questions. 

The paper by Ivezic08 showed that the observed variations in the
metallicity and velocity distributions of disk stars over the range $|Z|\sim1-3$
kpc are only mildly inconsistent with the traditional simple
decomposition into thin- and thick-disk components. However, they also found
that the rotational velocity and metallicity at $|Z|\sim1$ kpc, where the
contributions of the two components are similar, are uncorrelated. This
lack of correlation is in strong conflict with the traditional
decomposition. Instead, Ivezic08 modeled the observed distributions using
smooth shifts of the metallicity and velocity distributions that do not
change their shape. They argued that their ability to describe the
observations using functions with {universal $|Z|$-independent shapes} has
fundamental implications for disk origin -- instead of two distinct
components with { different} formation and evolution histories, the data
could be interpreted with a single, albeit complex, disk. 

On the other hand, Ivezic08 also pointed out that stars from the solar
neighborhood, kinematically selected as thick-disk stars, have larger
$\alpha$-element abundances, at the same $[Fe/H]$, than do thin-disk stars
(e.g., Bensby et al. 2004; Fuhrmann 2004; Feltzing 2006; Reddy et al. 2006;
Ram\'{i}rez et al. 2007). In addition, the thick-disk stars, again selected
kinematically, appear older than the thin-disk stars (e.g., Bensby et al.
2004; Fuhrmann 2004). Ivezic08 concluded that measurements of
$\alpha$-element abundances for samples of distant stars extending to
several kpc from the midplane (as opposed to local samples) could resolve
difficulties with traditional thin-thick disk decomposition when applied to
their data. The means to obtain such a dataset was recently produced by
\cite{Lee2011segue}, who showed that the $[\alpha/Fe]$ ratio can be estimated
using the comparatively low-resolution SDSS spectra -- for stars with
temperatures in the range 4500~K to 7000~K and sufficient signal-to-noise,
$[\alpha/Fe]$ estimates can be obtained with errors below 0.1 dex.

\subsection{The Holy Grail for Thin-Thick Disk Decomposition: $[\alpha/Fe]$} 

\begin{figure}%
\centerline{\psfig{figure=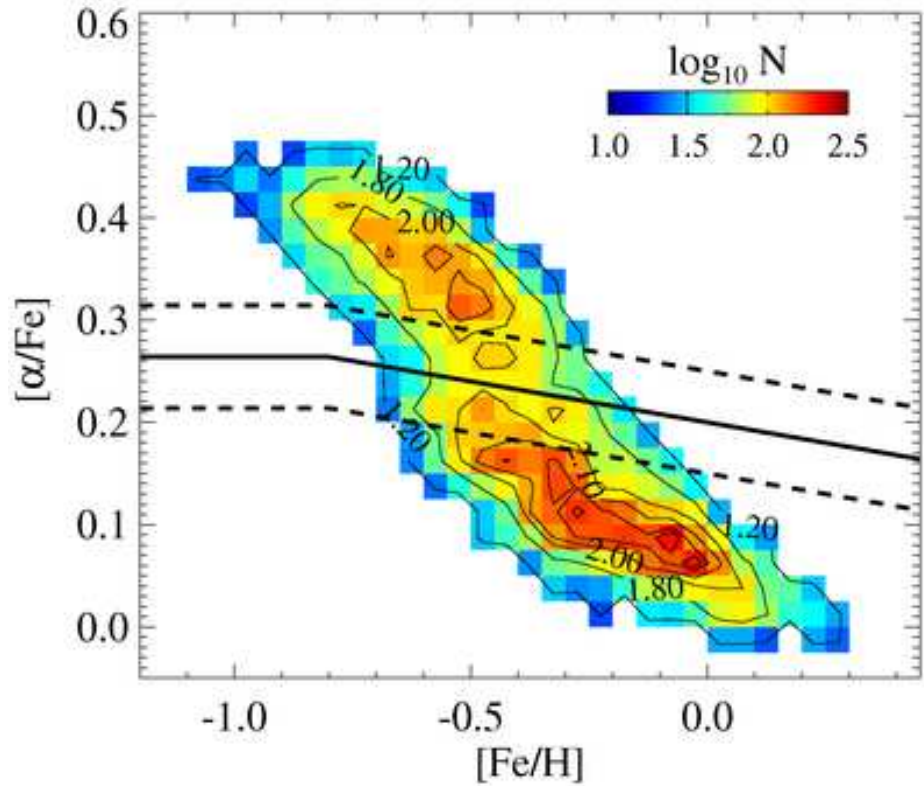,height=25pc}}
\caption{The $[\alpha/Fe]$ vs. $[Fe/H]$ distribution of G-type dwarfs within a few kpc 
from the Sun (Figure 2 from \citealt{Lee2011}). The number density
(arbitrarily normalized) is shown on a logarithmic scale according to the
legend, and by isodensity contours. Each pixel (0.025 dex in the
$[\alpha/Fe]$ direction and 0.05 dex in the $[Fe/H]$ direction) contains at
least 20 stars (with a median occupancy of 70 stars). The distribution of
disk stars in this diagram can be described by two components (thin disk
and thick disk, respectively) centered on ($[Fe/H],[\alpha/Fe]$) = ($-0.2,
+0.10$) and ($-0.6, +0.35$). The solid line is the fiducial for division
into likely thin- and thick-disk populations; note that a simple
$[\alpha/Fe]=0.24$ separation results in almost identical subsamples. The
dashed lines show the selection boundaries adopted by \cite{Lee2011}, which
exclude the central overlap region.}
\label{LeeFig2}
\end{figure}

\begin{figure}%
\centerline{\psfig{figure=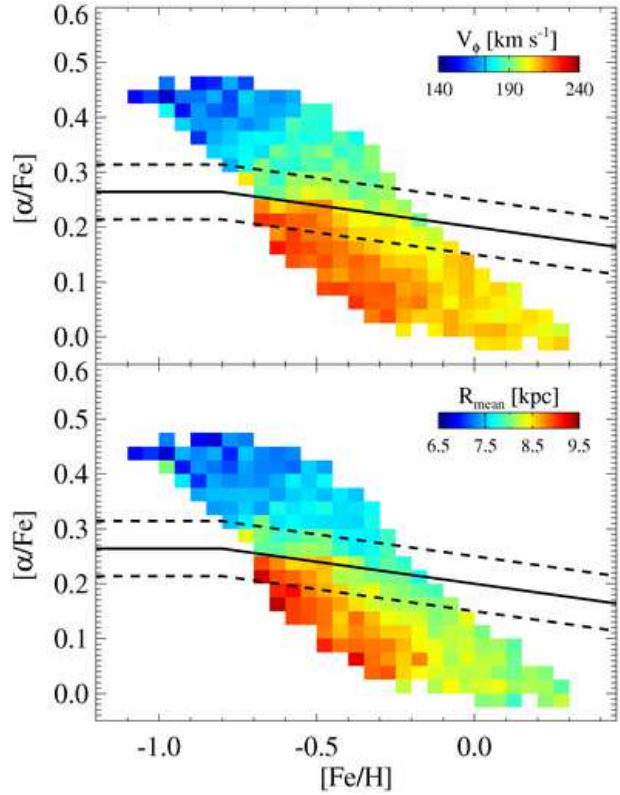,height=25pc}}
\caption{Distribution of mean rotational velocities ($v_\Phi$, top panel) and the orbital
radii ($R_{\rm mean}$, bottom panel) for the G-dwarf sample from Lee et al.
(2011b; their Figure 5) in the $[\alpha/Fe]$ vs. $[Fe/H]$ diagram
(3$\sigma$-clipped mean values). The orbital parameters are computed using an
analytic St\"{a}ckel-type gravitational potential from \cite{ChibaBeers2000}. The 
rotational velocity ($v_\Phi$) is defined in a { left-handed} coordinate system (the disk
rotation is $+220$ km s$^{-1}$). Note the rich structure present in both panels.}
\label{LeeFig5}
\end{figure}

\begin{figure}%
\centerline{\psfig{figure=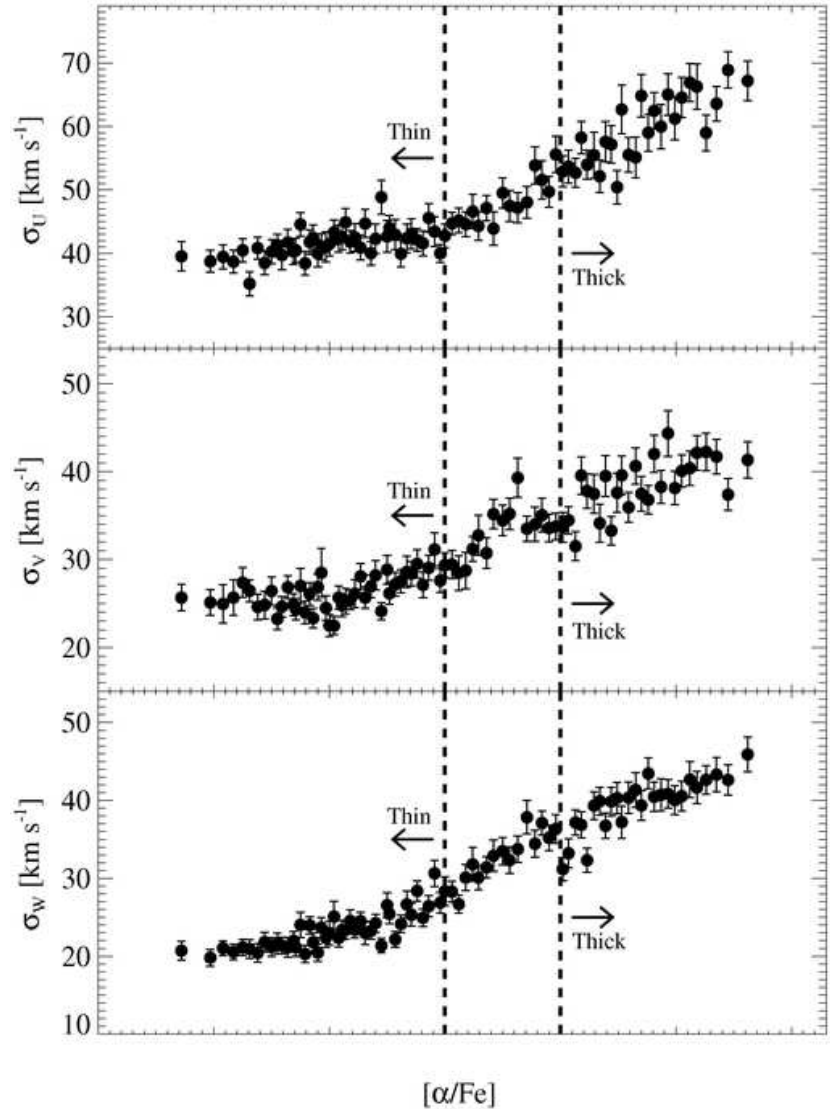,height=35pc}}
\vskip -0.3in
\caption{Variation of the three velocity dispersions with $[\alpha/Fe]$. 
 (Figure 3 from \citealt{Lee2011}). Each data point represents 200 
stars and the error bars are calculated by the bootstrap method. 
The vertical dashed lines at $[\alpha/Fe]= +0.2$ and $+0.3$ are added
to guide the eye, and roughly correspond to thin/thick disk separation.
The easily discernible increase of all three velocity dispersions with 
$[\alpha/Fe]$ provides kinematics-based support  for the chemical
$[\alpha/Fe]$-based separation of the two dominant disk components.  
}
\label{LeeFig3}
\end{figure}

\begin{figure}%
\centerline{\psfig{figure=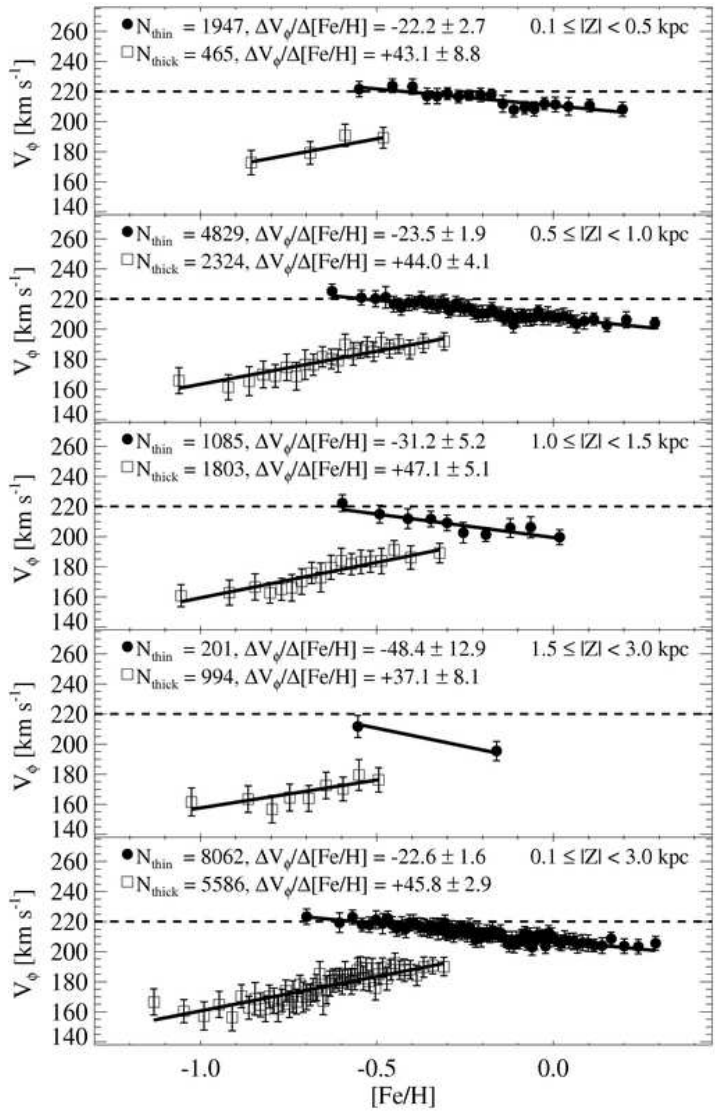,height=35pc}}
\vskip -0.3in
\caption{Variation of the mean rotational velocity of G-dwarf stars with metallicity for different slices in
distance from the Galactic plane (top four panels), for stars separated using
$[\alpha/Fe]$; thin-disk stars (black dots) and thick-disk stars (open squares)
are shown (Figure 7 from \citealt{Lee2011}). The rotational velocity ($v_\Phi$) is defined in a
 { left-handed} coordinate system (the disk rotation is $+220$ km s$^{-1}$).
Each dot represents a 3$\sigma$-clipped average of 100 stars. The bottom panel shows the results for
the full samples of stars considered. Estimates of the slopes and their errors
listed in the panels are computed for unbinned data.}
\label{LeeFig7}
\end{figure}

\citet[][hereafter L11]{Lee2011} analyzed a sample of $\sim$17,000 G-type dwarfs with 
$[\alpha/Fe]$ measurements based on the techniques of \cite{Lee2011segue}.
This dataset is the first massive sample of stars at distances of several
kpc with reasonably accurate distance estimates, measurements of all three
velocity components, measurements of both $[Fe/H]$ and $[\alpha/Fe]$, and
selected using well-understood and simple color and flux selection criteria
over a large area of sky. Thanks to these advantages, the L11 sample enabled a
number of far-reaching observational breakthroughs: 

\begin{enumerate}

\item The bimodal distribution of an unbiased sample of G-type dwarfs in the
$[\alpha/Fe]$ vs. $[Fe/H]$ diagram (see Figure \ref{LeeFig2}) strongly
motivates the separation of the sample by a simple $[\alpha/Fe]$ cut into two
subsamples that closely resemble traditional thin and thick disks in
their spatial distributions, $[Fe/H]$ distributions, and distributions
of their rotational velocity (see Figure 1 in L11).  

\item The low-$[\alpha/Fe]$, thin-disk subsample has an $[Fe/H]$
distribution that does not strongly vary with position within the
probed volume ($|Z|<$ 3 kpc and $7 < R / {\rm kpc} < 10$),  with 
a median value of $[Fe/H] \sim -0.2$. Similarly, the metallicity 
distribution for the high-$[\alpha/Fe]$, thick-disk subsample has
a median value of $[Fe/H] \sim -0.6$, without a strong spatial 
variation (see Figure 4 in L11).  

\item The rotational velocity component, $v_\Phi$, decreases linearly
with distance from the midplane, $|Z|$, with a gradient of
$d|v_\Phi|/d|Z|\sim -10$ km s$^{-1}$ kpc$^{-1}$ for both the thin- and
thick-disk subsamples (see Figure 8 in L11). The difference between the
mean values of $v_\Phi$ for the two subsamples of $\sim$30 km s$^{-1}$
(asymmetric drift) is independent of $|Z|$, and explains the discrepancy
between the $|Z|$ gradient of $-10$ km s$^{-1}$ kpc$^{-1}$ reported by L11,
and gradients about 2-3 times steeper reported for the full disk by earlier
studies (e.g., Ivezic08, Cassetii-Dinescu et al. 2011): as $|Z|$ increases
from the midplane to 2-3 kpc, the fraction of thick-disk stars increases
from $\sim$10\% to $>$90\%, and the observed gradient when all stars are
considered is affected by both the intrinsic gradient for each component,
and the velocity lag of thick-disk stars relative to thin-disk stars. 

\item The rotational velocity component does not exhibit a gradient
with respect to the radial coordinate, $R$, for thin-disk stars
($-0.1\pm0.6$ km s$^{-1}$ kpc$^{-1}$; a ``flat rotation curve''), and only
a small and marginally detected gradient for thick-disk stars ($-5.6\pm1.1$
km s$^{-1}$ kpc$^{-1}$).

\item The rotational velocity component and mean orbital radius are 
complex functions of the position in the $[\alpha/Fe]$ vs. $[Fe/H]$ diagram
(see Figure \ref{LeeFig5}). The rotational velocity component shows a
linear dependence on metallicity for both the thin- and thick-disk
$[\alpha/Fe]$-selected subsamples (see Figure \ref{LeeFig7}). The slopes of
these $v_\Phi$ vs. $[Fe/H]$ correlations have {opposite} signs,
$d|v_\Phi|/d[Fe/H] \sim -25$ km s$^{-1}$dex$^{-1}$ for the thin disk, and
$\sim 45$ km s$^{-1}$dex$^{-1}$ for the thick disk, and do not strongly
vary with distance from the midplane. These opposite gradients are
partially responsible for the lack of correlation between $v_\Phi$ and
$[Fe/H]$ at $|Z|\sim1$ kpc reported by Ivezic08 (for the full sample; the
other reason for the lack of correlation is systematic errors in the
photometric metallicity estimator, see Appendix in L11). 

\item Velocity dispersions for all three components of the local velocity
ellipsoid increase with $[\alpha/Fe]$ as smooth functions, and continuously
across the adopted thin/thick disk boundary (see Figure \ref{LeeFig3}).
Approximate values for the velocity dispersions ($\sigma_R, \sigma_Z,
\sigma_\Phi$) are (40, 25, 25) km s$^{-1}$ for the thin-disk subsample and
(60, 40, 40) km s$^{-1}$ for the thick-disk subsample, respectively (not
corrected for bias due to measurement errors; on average, about 10-15 km
s$^{-1}$ should be subtracted in quadrature). 

\item Orbital eccentricity distributions (model-dependent and determined using 
an analytic St\"{a}ckel-type gravitational potential from
\citealt{ChibaBeers2000}) are significantly different for the two
$[\alpha/Fe]$-selected subsamples (see Figure 10 in L11), and exhibit
strong variations with position and metallicity (see Figure 9 in L11).
Notably, the shapes of the eccentricity distributions for the thin- and
thick-disk populations are independent of distance from the plane, and
include only a minute fraction of stars with eccentricity above 0.6. 
\end{enumerate}

\begin{figure}%
\vskip -0.1in
\centerline{\psfig{figure=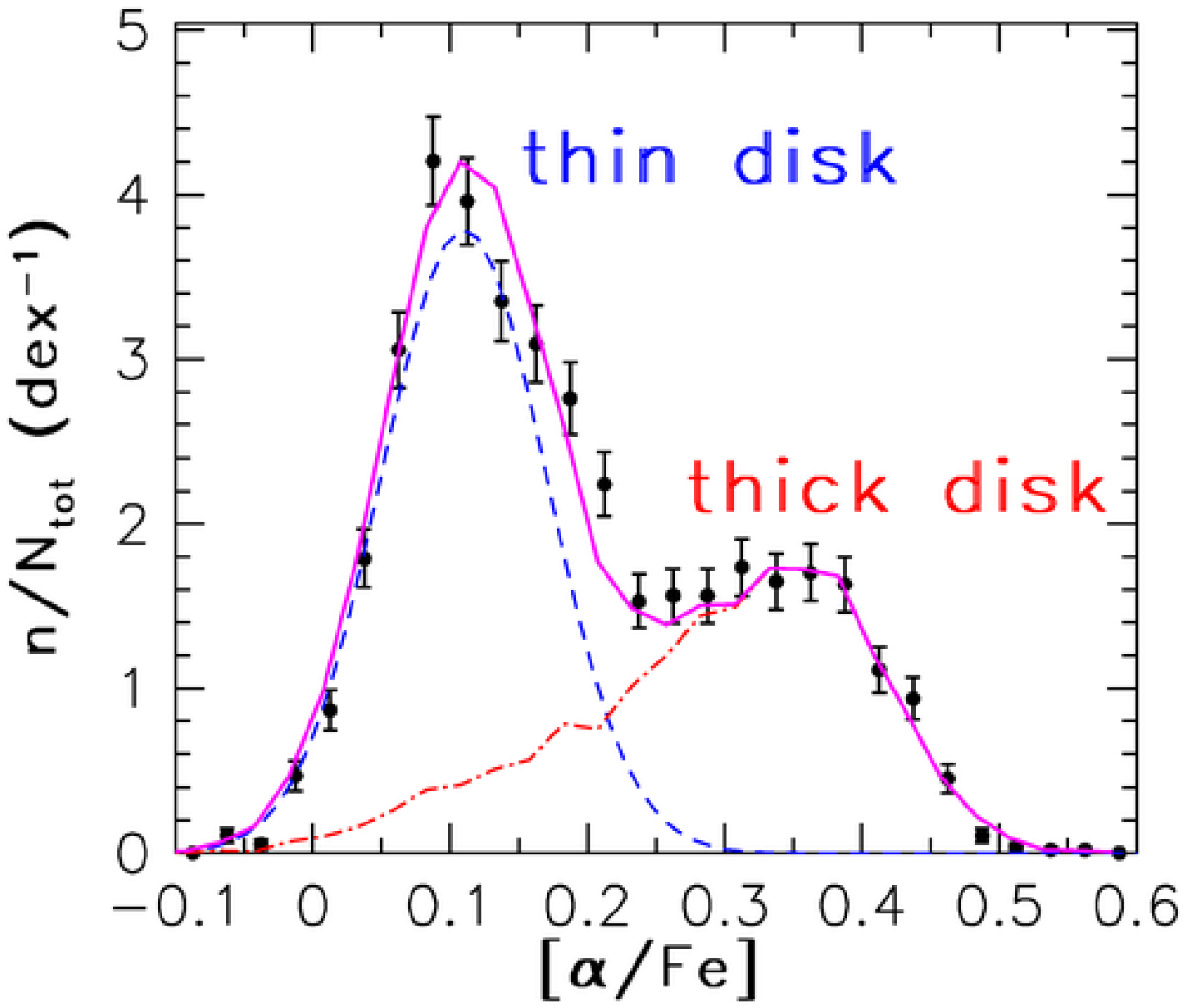,height=15pc},\psfig{figure=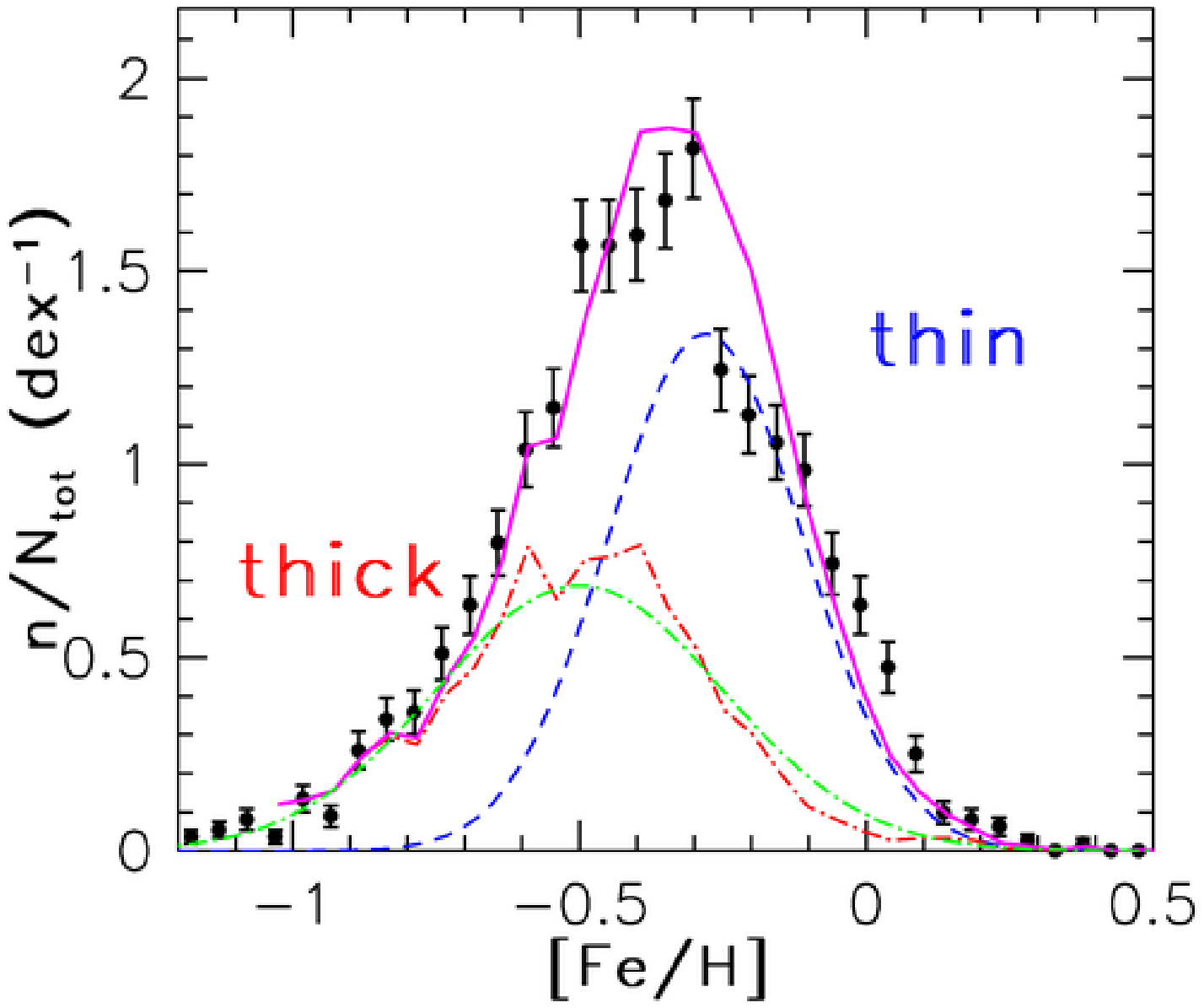,height=15pc},\psfig{figure=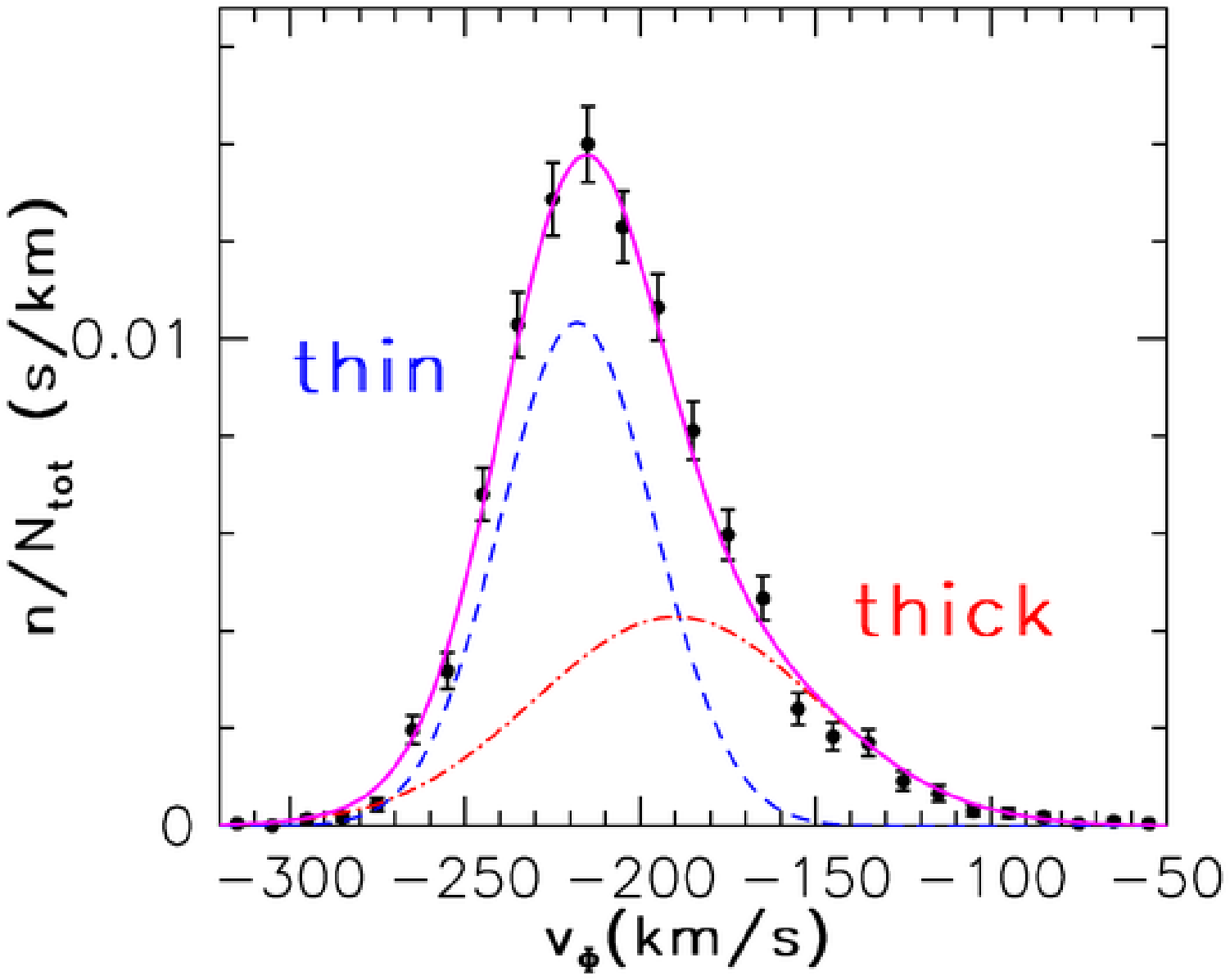,height=15pc}}
\vskip -1.2in
\caption{Tests of thin/thick-disk decomposition, using the sample of G-type
dwarfs from \cite{Lee2011}. The left panel shows the $[\alpha/Fe]$ distribution
for $\sim$2,300 stars in the fiducial bin $|Z|$=400-600 pc as symbols with
(Poissonian) error bars. The bimodality is easily seen. The observed
distribution can be modeled as the sum (shown by the magenta solid line) of two
components: the $[\alpha/Fe]$ distribution for $\sim$3,300 stars with $|Z|$=2-3
kpc shifted to lower values by 0.03 dex (red dot-dashed line) and a Gaussian
distribution, $N(0.11,0.06)$ (blue dashed line). The weights for the two components (0.43 
and 0.57, for the thick and thin component, respectively) are consistent with a double-exponential fit to star counts. 
The middle panel shows the $[Fe/H]$ distribution for the same stars from the fiducial 
$Z$=400-600 pc bin as symbols with error bars. Similar to the $[\alpha/Fe]$ distribution, 
it can be modeled as the sum (magenta solid line) of two components: the $[Fe/H]$
distribution for stars with $|Z|$=2-3 kpc shifted to higher values by 0.2 dex
(jagged red dot-dashed line) and $N(-0.28,0.17)$ (blue dashed line). The weights for
the two components (0.43 and 0.57) are the same as in the first panel. The
$[Fe/H]$ distribution for stars with $|Z|$=2-3 kpc is well described by
$N(-0.50,0.25)$ (after application of a 0.2 dex offset), shown as the smooth green dot-dashed line. 
The right panel shows the rotational velocity distribution for the same stars from the fiducial
$|Z|$=400-600 pc bin as symbols with error bars. It can be modeled as a linear
combination of two Gaussian distributions, $N(-218,22)$ and $N(-190,40)$,  
again using the same relative weights (and line styles) as in the first panel.}
\label{quick3}
\end{figure}

The behavior of the $[\alpha/Fe]$-selected subsamples of disk stars
strongly argues in favor of the traditional decomposition into two
(simpler) components. However, \cite{Lee2011} did not explicitly test
whether the counts of their two $[\alpha/Fe]$-selected subsamples are
consistent with the two best-fit additive exponential profiles obtained by
Juric08, nor did they test in detail the hypothesis that the variation of
the metallicity and rotational velocity distributions could be modeled
using two simple components weighted by the counts ratio. We have used data
from \cite{Lee2011} (kindly provided by Young Sun Lee) to perform these
tests here, as illustrated in Figure~\ref{quick3}. We confirm that
variations of the $[\alpha/Fe]$, $[Fe/H]$, and rotational velocity
distributions with $|Z|$ can indeed be interpreted as due to the interplay
of two simple components, whose relative strength variation with $|Z|$ is
consistent with the Juric08 results. In particular:

\begin{itemize}

\item The $[\alpha/Fe]$ distribution in the fiducial bin $|Z|$=400-600 pc is bimodal. 
It can be explained as a linear combination of the slightly modified
$[\alpha/Fe]$ distribution at $|Z|$=2-3 kpc (non-Gaussian, and presumably
dominated by the thick disk) and a Gaussian distribution with a mean of
$\langle [\alpha/Fe]\rangle$ = +0.11 and rms of 0.06 dex, and with weights
of 0.43 and 0.57, respectively (see the left panel in Figure~\ref{quick3}).
The only required modification of the $|Z|$=2-3 kpc $[\alpha/Fe]$
distribution is its shift towards lower $[\alpha/Fe]$ by 0.03 dex. The
weights for the two components are consistent with a double-exponential fit
to counts from Juric08, with the relative strength of the thick-disk
component increased from Juric08's value of 0.13 to 0.16 here (a $\sim
2\sigma$ change). 

\item The $[Fe/H]$ distribution in the $|Z|$=400-600 pc bin can be explained
as a linear combination of the slightly modified $[Fe/H]$ distribution at
$|Z|$=2-3 kpc (well described by a Gaussian) and a Gaussian distribution
with a mean of $\langle [Fe/H] \rangle = -0.28$ and rms of 0.17 dex, and
with the same weights as used for the $[\alpha/Fe]$ distribution (see the
middle panel in Figure~\ref{quick3}). The only required modification of the
$|Z|$=2-3 kpc $[Fe/H]$ distribution is its shift towards higher $[Fe/H]$ by
0.2 dex (uncertain to within 0.05-0.1 dex). 

\item The rotational velocity distribution in the $|Z|$=400-600 pc bin can be explained
as a linear combination of two Gaussian distributions (with $|v_\Phi|$
centered on 218 km s$^{-1}$ and 190 km s$^{-1}$, and with velocity
dispersions of 22 km s$^{-1}$ and 40 km s$^{-1}$, respectively), again with
the same weights as used for the $[\alpha/Fe]$ and $[Fe/H]$ distributions
(see the right panel in Figure~\ref{quick3}). When corrected for velocity
measurement errors (dominated by proper-motion errors), these dispersions
become 16 km s$^{-1}$ and 38 km s$^{-1}$, respectively.
\end{itemize}

The fact that all three distributions ($[\alpha/Fe]$, $[Fe/H]$, and
$v_\Phi$) in the $|Z|$=400-600 pc bin can be described as linear
combinations of the corresponding distributions in a distant bin dominated
by the thick-disk component and a best-fit thin-disk Gaussian, with { the
same weights} for all three cases that are consistent with the
double-exponential fit to the star counts, strongly supports the hypothesis
that the Milky Way disk comprises at least two distinct
components\footnote{Our analysis cannot exclude the possibility that the
disk structure is more complex than implied by the sum of only two simple
components. Indeed, Figure \ref{LeeFig3} shows that all three velocity
dispersions are smooth functions of the $[\alpha/Fe]$ ratio (and not step
functions, for example). This smoothness implies that an
$[\alpha/Fe]$-based disk decomposition into only two components is at best
a very good approximation, but definitely not the whole story. Most
recently (after this review was submitted), 
\cite{BRH2011} re-analyzed the same dataset and concluded that evidence for
the bimodal distribution of $[\alpha/Fe]$ all but disappears when selection effects are 
accounted for. The implication of  their result is that a continuous distribution of 
scale heights is a more appropriate model than a simple two-component model for 
describing SDSS data from L11.}.
The required shifts of $[\alpha/Fe]$ and $[Fe/H]$
distributions between the two $Z$ bins imply vertical gradients of $\sim0.015$
dex kpc$^{-1}$ and $\sim0.1$ dex kpc$^{-1}$ for the thick-disk component
(both with a relative uncertainty of about 30\%). 
Together with the rotational velocity gradient of $\sim10$ km s$^{-1}$
kpc$^{-1}$ (for both the thin- and thick-disk components) from \cite{Lee2011},
{ these vertical gradients represent strong constraints on models for
thick-disk formation.} 

We note that, out of six distributions (three quantities for the two
adopted disk components), the only strongly non-Gaussian distribution is
the thick-disk $[\alpha/Fe]$ distribution, that is, the $[\alpha/Fe]$
distribution for stars in the $|Z|$=2-3 kpc bin (dashed line in the left
panel of Figure~\ref{quick3}). Its skewness is due to the presence of about
15\% of the stars with $[\alpha/Fe]< +0.2$; their existence is puzzling.
According to \cite{Lee2011} and our own analysis of the kinematic and
metallicity behavior of stars with $[\alpha/Fe] < +0.2$, they represent the
thin-disk component. For example, in the $|Z|$=400-600 pc bin there are no
stars with $[\alpha/Fe] < +0.2$ that also have $[Fe/H] < -0.5$ or
$|v_\Phi|<140$ km s$^{-1}$, and stars with $[Fe/H]<-0.5$ and $|v_\Phi|<140$
km s$^{-1}$ have a median $[\alpha/Fe]$ of +0.40, with an rms of only 0.05
dex. However, the puzzle is that, according to the double-exponential fit
to star counts, the $|Z|$=2-3 kpc bin should contain only $\sim1$\% of
thin-disk stars, not 15\%. In addition, for stars with $|Z|$=2-3 kpc, the
subsample with $[\alpha/Fe] < +0.2$ has the same $[Fe/H]$ and $v_\Phi$
distributions as the subsample with $[\alpha/Fe] > +0.2$. Hence, it may be
that the skewed $[\alpha/Fe]$ distribution for stars in the $|Z|$=2-3 kpc
bin simply reflects a non-Gaussian measurement error distribution for the
method described in \cite{Lee2011segue}. It is noteworthy that stars from
the $|Z|$=2-3 kpc bin are on average about 3 mags fainter than stars from
the $|Z|$=400-600 pc bin ($r\sim18$ vs. $r\sim15$). Needless to say,
independent measurements of $[\alpha/Fe]$ for stars with $Z$=2-3 kpc would
provide valuable clues as to the proper interpretation. 

In summary, when disk stars are separated by a simple, well-motivated
$[\alpha/Fe]$ cut, the resulting subsamples display remarkably simple
spatial, kinematic, and metallicity distributions, consistent with the
traditional decomposition into thin- and thick-disk components. It is
likely that the differences in $[\alpha/Fe]$ reflect different
star-formation timescales (enrichment by Type Ia vs. Type II supernovae for
low and high $[\alpha/Fe]$ values over long and short timescales,
respectively; see \cite{BFL2004,Johnston2008}). Therefore, after detailed
analysis of full 7-D phase space, { SDSS data finally confirm that
$[\alpha/Fe]$ measurements provide the most robust decomposition of disk
stars into thin-disk and thick-disk components.} 

On the other hand, a few words of caution are due here. The main results
from L11 still need to be confirmed by independent datasets. It is somewhat
worrisome that the RAVE-based results from \cite{Burnett2011} for the disk
$[Fe/H]$ distribution differ from the L11 results. At $|Z|\sim0$,
the RAVE results are about 0.2 dex more metal rich (although we note that
the SDSS result for the median $[Fe/H]=-0.2$ at $|Z|=0$ is consistent with
the results from
\citealt{GCsurvey2004}), and the discrepancy increases to $\sim$0.3 dex at
$|Z|\sim2.5$ kpc. It is not clear yet whether the discrepant results reported by
the RAVE and SDSS surveys arise from differences in their adopted metallicity
scales, or are due to unaccounted selection effects in the RAVE analysis (see Section 6
in \citealt{Burnett2011}). Encouragingly, the spatial metallicity gradients at
$|Z|\sim1$ kpc, where thick-disk stars become more numerous than thin-disk
stars, are robustly detected and similar in both studies, $d[Fe/H]/d|Z| \sim
-0.2$ dex kpc$^{-1}$ (for stars from both components considered together). 
The median $[Fe/H]$ at $|Z|\sim1$ kpc reported by
\cite{Lee2011} is $-0.5$ dex, about 0.2 dex lower than reported by 
\cite{Burnett2011} using RAVE, and about 0.2 dex higher than reported
by Ivezic08 using photometric metallicities from the SDSS imaging survey. 
It remains to be seen how the $[\alpha/Fe]$ measurements from the SDSS and
RAVE surveys compare to each other; further study will presumably provide
illumination. 

The \cite{Burnett2011} study also reports age determination for RAVE stars
(based on stellar models), with typical uncertainties of about a factor of two (see
their Figure 7). They detect a remarkable age gradient between the Galactic
midplane and $|Z|\sim2$ kpc (see their Figures 16 and 17), which is at least
qualitatively consistent with the variation of the $g-r$ color of turnoff stars
seen by SDSS, and the velocity dispersion-age correlations for local disk stars
from \citet{GCsurvey2004}, \citet{Helio04}, and \citet{West08}. They also detect
a complex variation of metallicity distribution with stellar age (see their
Figure 18). In particular, the oldest stars ($> 8-9$ Gyr) are predominantly
low-metallicity ($[Fe/H]<-0.5$). These age data represent a valuable addition to
the L11 results. Nevertheless, determining age for individual stars is exceedingly
difficult \citep{PontEyer2004, Soderblom2010} and one needs to consider all the
caveats discussed by Burnett et al. at the end of their Section 7. Given
these difficulties, it seems best to proceed with caution.  There would be
clear advantage in, at the very least, obtaining age estimates for the SDSS
sample (using different mehtodologies than used for the RAVE data), and
making a more direct comparison based on this information.

Last but not least, we note that the observational material for studying the
bulge of the Milky Way has also significantly improved during the last decade.
\cite{Clarkson2008} used HST to detect proper motions for over 15,000 bulge
stars and ``dissected'' the kinematic properties of the bulge as a function of
distance along the line of sight. A radial-velocity survey of bulge stars
\citep[BRAVA;][]{Rich2011} obtained data for 10,000 red giants in the Southern
Galactic bulge, and found clear departures from solid-body rotation that are
consistent with an edge-on bar. \cite{RW2009} and \cite{RWS2009} reported
radial-velocity and metallicity measurements for over 3,000 bulge stars, and now
even $[\alpha/Fe]$ measurements are available for large samples of stars. For
example, \cite{Gonzales2011} determined $[\alpha/Fe]$ for 650 red-giant stars
using $R\sim22000$ spectroscopy, and found support for two bulge components in
the observed metallicity and $[\alpha/Fe]$ distributions, reminiscent of the
disk separation into thin- and thick-disk components. They argue that the
chemical similarity of the low-metallicity bulge component and the thick disk
hints for rapid, early formation for both structures. It is likely that such
observational progress will lead to additional studies that simultaneously
consider all of the major structural components of the Galaxy.

%
%

\subsection{Comparisons of Observations with Disk Formation Models} 

Despite the past three decades of thick-disk studies, there is still no
consensus on models for its formation and evolution (the thick disk is not
unique to the Milky Way; for a review of thick disks in other galaxies, see
\citealt{KruitFreeman2011}). The proposed scenarios can be broadly divided into
two groups: violent origin, such as heating of an existing thin disk due to
mergers, and secular evolution, such as heating due to scattering off molecular
clouds and spiral arms (see L11 for a detailed discussion and references). In
the first set of scenarios, the fraction of thick-disk stars accreted from
merged galaxies remains an important and still unconstrained parameter, and
further complexity arises from the possibility that some stars may have formed
{in situ}, when star formation is triggered in mergers of gas-rich galaxies
\cite[][and references therein]{Brook2007}. In the second set of scenarios, the
main modeling difficulty is the lack of detailed knowledge about the relative
importance of various scattering mechanisms. Over the last decade, the 
radial-migration mechanism (\citealt{SellwoodBinney2002}; \citealt{Roskar2008a};
\citealt{SchoenrichBinney2009b}; \citealt{MinchevFamaey2010}) has been developed
as an attractive secular scenario. Due to various computational and other
difficulties, numerical models that combine the main features of the violent and
secular scenarios are few. 

The recent observational material contains rich information for model
testing, and is beginning to rule out some models. Modern data include
simultaneous measurements of many observables for large numbers of stars,
and enable qualitatively new approaches to tests of disk-formation models.
The more observables that are measured, the more powerful these tests
become, because the data can be ``sliced'' along multiple axes in a variety
of ways while maintaining small statistical errors due to the large sample
sizes. For example, the two eccentricity distributions for
$[\alpha/Fe]$-selected subsamples are much more powerful model
discriminators than the eccentricity distribution for all stars lumped
together. On the other hand, the complexity of such tests can be
formidable -- even a minimalistic selection of observables, such as
coordinates $R$ and $|Z|$, chemical parameters $[Fe/H]$ and $[\alpha/Fe]$,
and the essential kinematic parameters, rotational velocity and orbital
eccentricity, span a six-dimensional space. The basic model vs. data
comparisons for testing thick-disk formation and evolution scenarios
include:

\begin{enumerate}

\item Comparison of the observed distribution of stars in the $[\alpha/Fe]$ vs. 
$[Fe/H]$ diagram, as a function of the position in the Galaxy (e.g., Can models
reproduce the bimodal distribution seen in Figure \ref{LeeFig2}? Does the
fraction of the sample in the high-$[\alpha/Fe]$ component increase with the
distance from the midplane as observed?). 

\item For subsamples defined using $[\alpha/Fe]$, comparison of
the shapes of their metallicity and kinematic distributions (e.g., Can models
reproduce the $[Fe/H]$ distributions seen in Figure 4 from L11 and in Figure
\ref{quick3}, or the eccentricity distributions seen in their Figure 10?). 

\item For subsamples defined using $[\alpha/Fe]$, comparison of the variations 
of their number density and low-order statistics for metallicity and
kinematic distributions (e.g., $\langle v_{\Phi} \rangle$, velocity dispersions,
mean/mode/median eccentricity) with position in the Galaxy (e.g., Can models
reproduce the spatial gradients of the $\langle v_{\Phi} \rangle$ seen in Figure 8 from L11,
or the spatial gradients of the mean eccentricity from their Figure 9?). 

\item Comparison of high-order correlations between the observables, such as the 
complex variation of the mean rotational velocity with position in the
$[\alpha/Fe]$ vs. $[Fe/H]$ diagram (see Figures \ref{LeeFig5} and
\ref{LeeFig7}), or the variation of the orbital eccentricity with metallicity
(see Figure 9 in L11). 

\end{enumerate}

A few of the above tests have already been performed. In a strict
statistical sense, all the proposed models can be outright rejected because
the observed distributions of various parameters have very low statistical
noise, and the models are not sufficiently fine tuned (yet) to reproduce
them (e.g., none of model eccentricity distributions comes even close to
passing the Kolmogorov-Smirnov test). For this reason, most of the model vs.
data comparisons are still qualitative, and only gross inconsistencies can
be used to reject certain scenarios. 

Beginning with \cite{Sales2009}, a number of recent papers have used the
shape of the derived orbital eccentricity distribution as a means to
compare models to data from the SDSS and RAVE surveys \citep{Dierickx2010,
DiMatteo2011, CasettiDinescu2011, Wilson2011, Loebman2011,Lee2011}. We note
that orbital eccentricity is derived from observations in a model-dependent
way (a gravitational potential must be assumed), and different assumptions
may lead to systematic differences between the observed and predicted
distributions. Another detail to keep in mind is that stars with small
rotational velocities are often excluded to minimize the contamination of
disk samples by halo stars. However, disk stars with very high orbital
eccentricity are also excluded by the same cut, and their exclusion may
lead to unjustified model rejection. Although there are detailed
differences in the eccentricity distributions derived from data, the mode
of the distributions for stars at about 1-2 kpc from the midplane is
typically in the range 0.2--0.3, and the fraction of stars with
eccentricities larger than 0.8 are below a percent or so (unfortunately,
none of recent papers listed above show cumulative distributions, nor
directly compare with various data-based distributions). 

In most of the recent studies, four published simulations of thick disks
formed by (a) accretion from disrupted satellites, (b) heating of a
pre-existing thin disk by a minor merger, (c) radial migration and (d)
gas-rich mergers (see Sales et al. for references), are confronted with
data. The model predictions for eccentricity distributions are nicely
summarized in Figure 3 from \cite{Wilson2011} and Figure 17 from
\cite{CasettiDinescu2011}. Scenario (a) produces an eccentricity
distribution with a mode at $\sim$0.5, and scenario (b) predicts a bimodal
eccentricity distribution that includes too many stars ($\sim$10\%) with
eccentricities above 0.8 (see Figure 3 in Sales et al. and Figure 10 in
L11, but note that stars with eccentricity larger than 0.8 exist in the
sample from \citealt{CasettiDinescu2011}). In addition, scenario (b) does
not exhibit the characteristic change of slope in the log(counts) vs. $|Z|$
plot (see Figure 1 in Sales et al.). These discrepancies are the main
reasons for the growing consensus that the gas-rich mergers and
radial-migration scenarios are in best agreement (more precisely, least
disagreement) with the present data. 

\begin{figure}%
\centerline{\psfig{figure=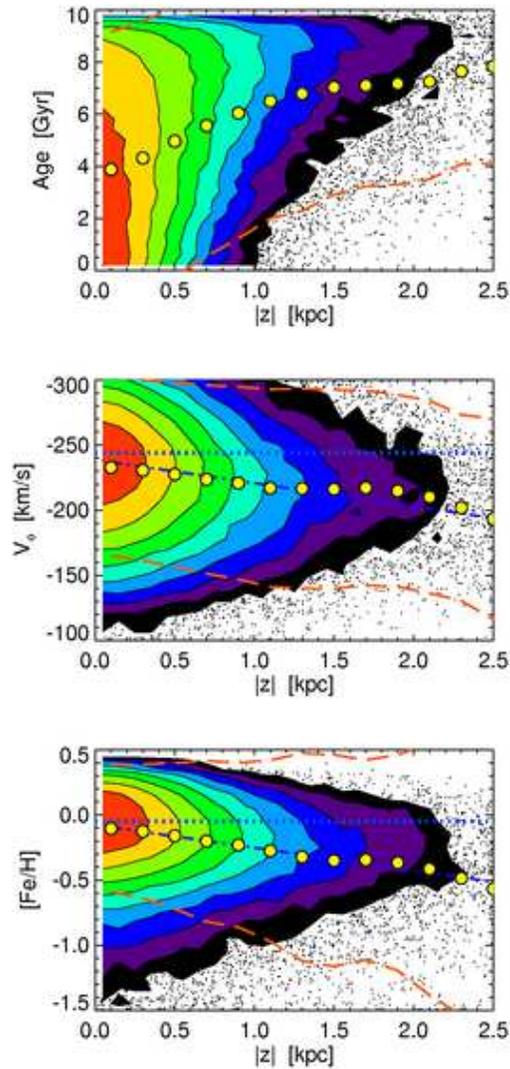,height=35pc}}
\caption{Predictions of the radial-migration model from 
\cite{Roskar2008a} for the variation of stellar age, rotational 
velocity, and metallicity with distance from the Galactic plane
for stars in the solar cylinder (Figure 8 from \citealt{Loebman2011}). 	   
The simulated distributions are represented by color-coded contours 
(low to medium to high: black to green to red) in the regions of high 
density, and as individual points otherwise. The large symbols show the 
means for the $|Z|$ bins, and the dashed lines show a 2$\sigma$ envelope.
The gradients seen in the bottom two panels are consistent with the
SDSS-based results.}
\label{LoebmanFig7}
\end{figure}

\cite{Loebman2011} performed a number of the data vs. model tests listed above, in the limited 
context of the radial-migration models developed by \cite{Roskar2008a,
Roskar2008b}. They demonstrated that the overall features seen in the SDSS
data, such as the gradients of metallicity and rotational velocity with
distance from the midplane (see Figure~\ref{LoebmanFig7}), as well as the
gradients of rotational velocity with metallicity (see their Figure 15),
and the complex structure seen for the mean rotational velocity in the
$[\alpha/Fe]$ vs. $[Fe/H]$ diagram (their Figure 14), are qualitatively
reproduced by the models (at a detailed quantitative level there is room
for improvement). We note an important implication of those models -- 
{$[\alpha/Fe]$ should be an excellent proxy for age.} Using a different
numerical implementation of the radial-migration scenario,
\citet{SchoenrichBinney2009a} and \citet{SchoenrichBinney2009b,RS2011} demonstrated
good agreement with the local solar neighborhood data from the
Geneva-Copenhagen survey \citep{GCsurvey2004}. 

These model successes hint that the thick disk may be a ubiquitous Galactic
feature generated by stellar migration (though note that a similar analysis
has not yet been carried out with the gas-rich merger models). However,
although these models at least qualitatively reproduce much of the
complex behavior seen in the data, radial migration cannot be the full
story -- there exist counter-rotating disks observed in some galaxies
\citep{YoachimJD2008}, thick disks are less prominent in high-mass galaxies
\citep{YoachimJD2006}, and remnants of merged galaxies are directly
observed in the Milky Way (see the right column in Figure~\ref{TIf26} and
the discussion in \S\ref{Sec:substructure} below).

\subsection{A Summary of Recent Disk Studies} 

To summarize, given the new SDSS, RAVE, and other data, there is no doubt
that the spatial and kinematic behavior of disk stars greatly varies as a
function of their chemical composition, parametrized by the position in the
$[\alpha/Fe]$ and $[Fe/H]$ diagram. While quantitative details still differ
somewhat between different analysis methods, and between the SDSS and RAVE
datasets, robust conclusions are that the high-$[\alpha/Fe]$ subsample has
all the characteristics traditionally assigned to the thick disk: larger
scale height, lower $[Fe/H]$, a rotational velocity lag, and larger
dispersions for all three velocity components, when compared to the
low-$[\alpha/Fe]$ subsample. There is mounting evidence that the ages of
these stars are higher than those in the low-$[\alpha/Fe]$ subsample, and
similar to the age of Galaxy, although the interpretation of age data is
much more prone to systematics than chemical and kinematic data. 


{ Despite this tremendous observational progress, there is still no
consensus on theories for the origin of thick disk.} The two main contenders
remain gas-rich mergers and radial-migration scenarios, while the accretion 
and disk heating scenarios appear to be in conflict with the data. Nevertheless, no generic
model/scenario should be fully rejected yet, since detailed comparisons with
data have only begun and the input model parameter space has not been fully explored.
Assuming that SDSS measurements reported in L11 survive further scrutiny
(e.g., when compared to RAVE and other datasets), modelers will be kept busy
for some time trying to explain the rich observational material collected
over the last few years.

\section{THE MILKY WAY HALO}

Studies of the Galactic halo provide unique insights on the formation history of
the Milky Way, and for the galaxy formation process in general, because dynamical
timescales are much longer than for disk stars and thus the ``memory of past events
lasts longer''\citep[e.g.,][]{jhb96,may02}. The last decade has seen tremendous
progress in both observations and simulations of the Milky Way halo. For
example, \cite{WMcG1996} pointed out that there were only nine RR Lyrae stars
discovered at Galactocentric distances larger than 30 kpc at that time. With the
advent of SDSS, 2MASS, QUEST, and other surveys, there are now many hundreds of
RR Lyrae stars \citep{vz06,Sesar2010} and thousands of BHB
stars \citep{Brown2010,Sirko04,Xue08} detected all the way to $\sim$100 kpc.
The 2MASS point source catalog has provided an all-sky view of the distribution
of M giants beyond 30 kpc \citep{Majewski2003}, and a large sample of
carbon-rich giants have been observed at similar distances \citep{Ibata2001}. With SDSS
data, it is now possible to study the halo within $\sim$10-20 kpc using tens of
millions of main-sequence stars \citep[][Juric08, Ivezic08, Bond10]{new02,new07, Belo2007c,
Bell2008}. Concurrently, models for the formation and evolution of stellar halos
have grown increasingly more sophisticated and predictive \citep{Law2005,
Bullock2005,Johnston2008,Helmi08, Stark2009, Ghigna2000,Bullock2001,
Springel2008,Font2011}. For example, contemporary simulations of galaxy formation predict
that stellar halos of Milky Way-type galaxies are assembled from inside out,
with the majority of the mass (50\%-80\%) coming from several massive
(10$^8$-10$^{10}$ \Mdot) satellites that have merged more than 9 Gyr ago, while
the remaining mass comes from lower mass satellites accreted in the past 5-9 Gyr
\citep{Bullock2005,DeLuciaHelmi2008,Font2011}. 

The new data collected over the last decade led to significantly improved
quantitative understanding of the spatial distribution, kinematics, and
metallicity distribution of halo stars. We first review results for the
relatively nearby halo within $\sim$20 kpc, probed in situ\footnote{Hereafter we use { in situ} to refer to 
measurements of stars located where their stellar population dominates, 
as opposed to ``extrapolated properties'' based on local samples.}
with main-sequence stars, then summarize observations out to $\sim$100 kpc with
various other more luminous tracers, and finish with a discussion of stellar
streams and other halo substructures.

\subsection{The Smooth Halo Behavior as Probed by Main-Sequence Stars} 

Despite the presence of abundant halo substructure (see below), it is possible to describe
the distribution of halo stars within $R \sim$20 kpc at an impressive level of fidelity
using a simple smooth, oblate, and cylindrically symmetric power-law model.
The Juric08 study used SDSS data for $\sim$50 million stars, together with a photometric
parallax method, to estimate their distances (see \S\ref{ppmethod}), and found
that the local stellar halo can be modeled as 
\begin{equation}
\label{eq:halorho}
            n(R,Z) = n_o f_H \left({  R_\odot^2   \over R^2 +  (Z/q_H)^2}\right)^{n \over 2},
\end{equation} 
where $n(R,Z)$ is the number of stars per unit volume (``number density''), as a
function of the cylindrical coordinates $R$ and $Z$, $n_o$ is the local (at the
solar position) number density of all stars, $f_H$ is the local fraction of halo
stars, and $q_H$ parametrizes deviations from spherical symmetry ($q_H<1$ for an oblate
halo). After masking regions with obvious localized overdensities, Juric08 obtained
the following best-fit parameters: $f_H=0.005$, $q_H=0.64$, and $n=2.8$.
Examples of the observed $n(R,Z)$, the best-fit model, and the fit residuals from
the Juric08 study are shown in Figure~\ref{TIf26} (note that the disk component is also
shown; see also the bottom panel in Figure~\ref{TIf15}). It is important to
remember that the dataset used by Juric08 does not extend beyond $R\sim20$ kpc and
$|Z|\sim10$ kpc. Indeed, additional data suggest that the Juric08 single power-law
halo cannot be extrapolated beyond 20 kpc (see below). 

Recently, Bonaca et al. (2012) examined whether eq.~\ref{eq:halorho}  is an
appropriate model for the distribution of main-sequence turn-off halo stars observed by
the Eight Data Release of the SDSS (DR8, \citealt{Aihara2011}). DR8 covers almost a factor
of two more area than analyzed by Juric08, including an order of
magnitude greater coverage of the Southern Galactic hemisphere. After
accounting for known overdensities and streams (Virgo, Hercules-Aquilae, Sagittarius;
see \S~\ref{Sec:substructure}), Bonaca et al. (2012, to be submitted) find no evidence for 
any residual triaxiality of the stellar halo. To the extent that can be
probed with SDSS imaging data, the halo within 10-20 kpc from the Galactic
center remains well described by an oblate ellipsoid (for a discussion of more distant 
parts of the halo, see \S~\ref{Sec:OuterHalo}. 

Although the SDSS spectroscopic survey has provided metallicity
measurements for a large number of stars, spectroscopic estimates of
$[Fe/H]$ are available for $<1$\% of the stars used by Juric08. To provide
a panoramic map of the $[Fe/H]$ distribution for halo stars, Ivezic08
utilized a photometric metallicity method (see
\S\ref{pmetmethod}). Their $[Fe/H]$ map, shown in Figure~\ref{FeHmap},
demonstrates that the median metallicity of halo stars is essentially invariant
within the probed volume; Ivezic08 determined an upper limit for its spatial gradient
of 0.005 dex kpc$^{-1}$ within $|Z|<10$ kpc. The local halo $[Fe/H]$ distribution is
well described by a Gaussian centered on $[Fe/H]=-1.46$, with an rms of 0.30
dex. 

\begin{figure}%
\centerline{\psfig{figure=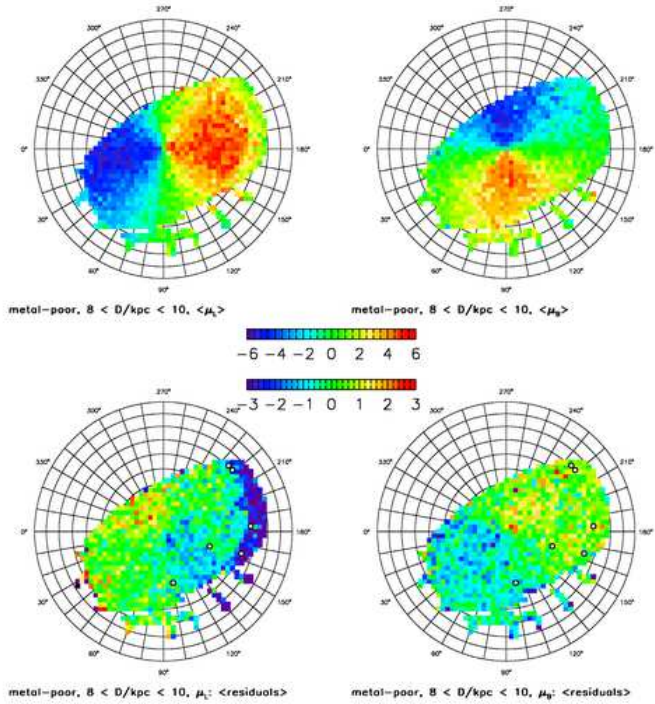,height=35pc}}
\vskip -0.9in 
\caption{Figure 21 from \cite{Bond2010}. 
Distribution of the median longitudinal proper motion in a Lambert
projection of the North Galactic cap for low-metallicity
(spectroscopic $[Fe/H] < -1.1$), blue ($0.2 < g-r < 0.4$) stars,
with distances in the range 8-10 kpc. The top two panels show
the median longitudinal (left) and latitudinal (right) proper motions,
and the two bottom panels show the median difference between the
observed and model-predicted values. The maps are color-coded
according to the legends in the middle (mas yr$^{-1}$); note that the
bottom scale has a harder stretch to emphasize structure in the
residual maps). In the bottom panels, the white symbols show the 
positions of the six northern cold substructures (see \S\ref{Sec:substructure}) 
identified by \cite{Schlaufman09}.}
\label{TIIIf21}
\end{figure}

\begin{figure}%
\centerline{\psfig{figure=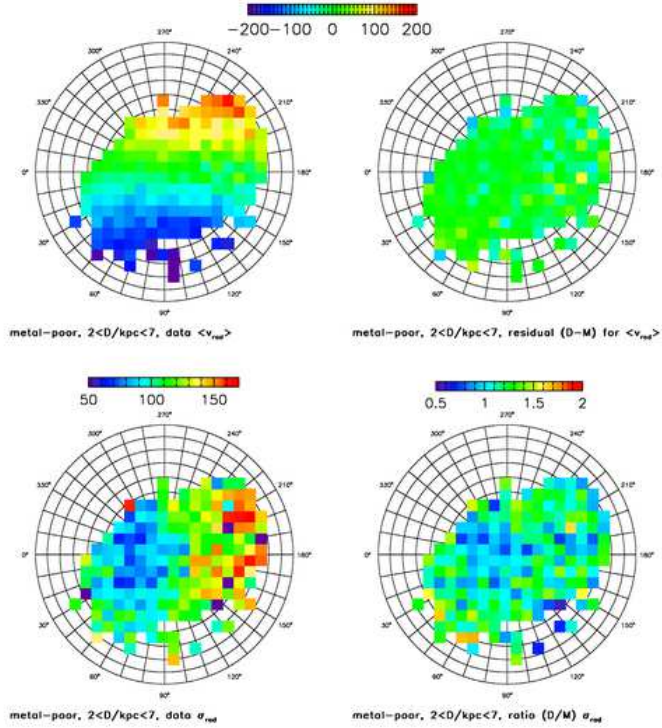,height=35pc}}
\vskip -0.9in 
\caption{Figure 16 from \cite{Bond2010}.
Comparison of medians and dispersions for the measured and modeled
radial velocities of 20,000 blue ($0.2<g-r<0.4$) halo stars
(spectroscopic $[Fe/H] < -1.1$) at distances $D = 2 - 7$ kpc, and 
$b > 20^\circ$. The top-left panel shows the median measured radial
velocity in each pixel, color-coded according to the legend shown at
the top (units are km s$^{-1}$). The top-right panel shows the
difference between this map and an analogous, visually similar, map based on
model-generated values of radial velocity, using the same scale as in
the top-left panel. The bottom-left panel shows the dispersion of
measured radial velocities, color-coded according to the legend above
it. The bottom-right panel shows the ratio of this map and an
analogous, visually similar, map based on model-generated values of radial 
velocity, color-coded according to the legend above it. 
}
\label{TIIIf16}
\end{figure}

\begin{figure}%
\vskip -4.2in 
\centerline{\psfig{figure=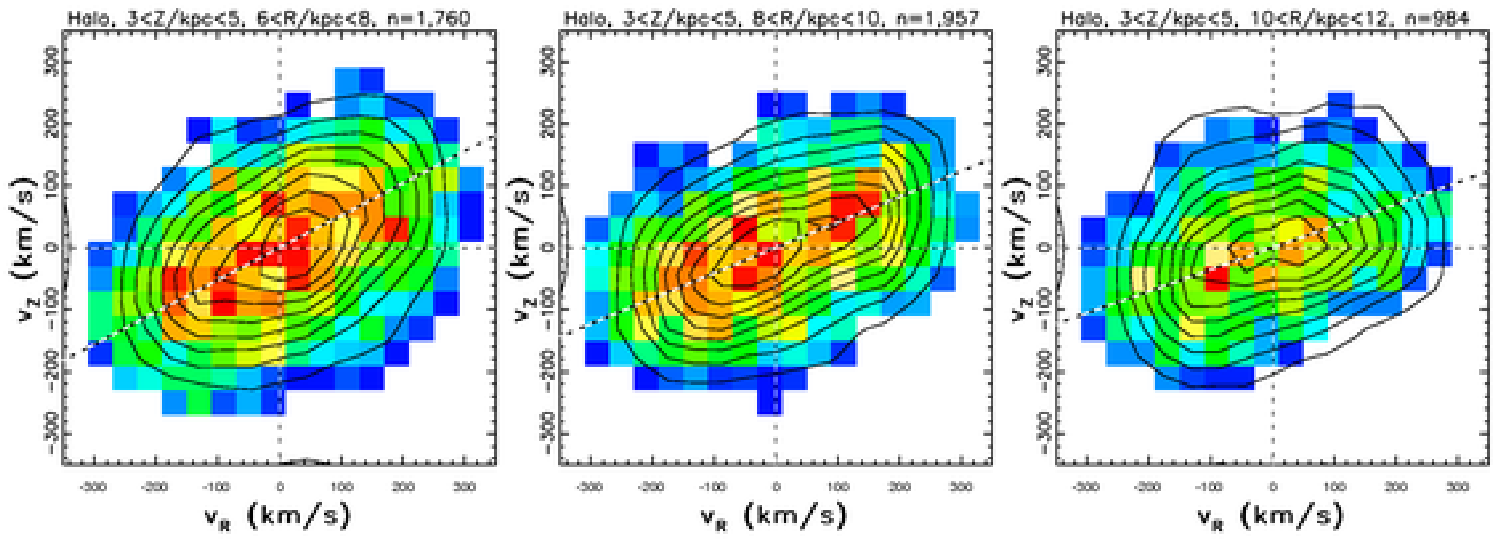,height=45pc}}
\vskip -0.5in 
\caption{Figure 13 from \cite{Bond2010}. The two-dimensional $v_Z$
vs. $v_R$ projections of the velocity distribution for three
subsamples of candidate  halo stars selected using spectroscopic
metallicity ($-3 < [Fe/H] < -1.1$), with $3 < |Z|/{\rm kpc}  < 5$, and
$6 < R/{\rm kpc} < 8$ (left), $8 < R/{\rm kpc} < 10$ (middle), and $10
< R/{\rm kpc} < 12$ (right).  These $R-Z$ boundaries are illustrated in 
Figure~\ref{FeHmap}.   
The distributions are shown using linearly spaced contours, and with a
color-coded map showing smoothed counts in pixels (low to high from
blue to red). The measurement errors are typically 60 km s$^{-1}$, and the 
dashed lines show the median direction toward the Galactic center.
Note the strong evidence for a velocity-ellipsoid tilt,  and the
variation of the tilt with $R$, so that the ellipsoid always points
towards the Galactic center.}
\label{TIIIf13}
\end{figure}

In the third paper of the Milky Way tomography series, Bond10 used the
large database of SDSS-POSS proper motions (see \S\ref{sdssposs}), and
radial-velocity measurements from the SDSS spectroscopic survey (see
\S\ref{sdss}), to quantify the kinematic behavior of halo stars. Similarly
to the behavior of counts and metallicity, Bond10 found that halo
kinematics also admit a simple model description. The very complex behavior
of measured proper motions (see Figure~\ref{TIIIf21}) and radial velocities
(see Figure~\ref{TIIIf16}) on the sky can be explained with a simple
triaxial velocity ellipsoid that is invariant in spherical coordinates,
with $\sigma_r$=141 km s$^{-1}$, $\sigma_\Phi$=85 km s$^{-1}$,
$\sigma_\theta$=75 km s$^{-1}$, and their uncertainties of $\sim$5 km
s$^{-1}$. For example, the substantial variation in the dispersion of
measured radial velocities {across the sky} seen in the bottom left panel
in Figure~\ref{TIIIf16} is due to the change of the orientation of velocity
ellipsoid {with respect to the line of sight} (see Figure~\ref{TIIIf13}),
rather than some localized substructure. A similar triaxial velocity
ellipsoid was measured by \cite{Smith2009MNRAS}, using more robust proper
motions based on only SDSS astrometry (as opposed to the SDSS-POSS dataset
used by Bond10), although in only a single direction on the sky. This
remarkable alignment of the halo velocity ellipsoid with spherical
coordinates (halo stars ``know'' where the Galactic center is -- see
Figure~\ref{TIIIf13}!) is also supported by independent data from the RAVE
survey \citep{Siebert08}, and represents a strong constraint on the shape
of gravitational potential -- the potential must be close to spherically
symmetric within $\sim$20 kpc from the Galactic center 
\citep[][and references therein]{Smith2009ApJ}. The
spherical symmetry of the gravitational potential is also invoked as an
explanation for the lack of precession of the orbital plane of debris of the
Sagittarius dwarf spheroidal (see below) by \cite{Fellhauer2006}, which is
implied {if} the observed bifurcation in the distribution of debris is due
to multiple (young and old) streams. On the other hand, \cite{Helmi2004}
concluded that the dark matter halo has a prolate shape, based on modeling the
dynamics of the leading (old) stream. 

It is noteworthy that the difference of 10 km s$^{-1}$ between
$\sigma_\Phi$ and $\sigma_\theta$ measured by Bond10 is only marginally
detected (less than 2$\sigma$ significance if the measurement errors are
$\sim$5 km s$^{-1}$, as claimed by Bond10). An even smaller difference was
measured by \cite{Smith2009MNRAS}, using proper motions based on only SDSS
astrometry ($\sigma_\Phi$=82 km s$^{-1}$ and $\sigma_\theta$=77 km
s$^{-1}$), with quoted errors of $\sim$2 km s$^{-1}$ (which are likely to
be underestimated because distance errors are not taken into account; both
Bond10 and Smith et al. used the photometric parallax relation from
Ivezic08). The significance of the difference between $\sigma_\Phi$ and
$\sigma_\theta$ is important, because if $\sigma_\Phi = \sigma_\theta$, then
the halo stellar density distribution should not be oblate, as measured by
Juric08, but spherical instead \citep{Smith2009ApJ}. Indeed, using
kinematic constraints based on their measurements of $\sigma_\Phi$ and
$\sigma_\theta$, \cite{Smith2009MNRAS} obtained an implied stellar halo
flattening parameter of $q_H=0.98$, in puzzling disagreement with the
value of $q_H=0.64$ measured {in situ} by Juric08. Nevertheless, Smith et al.
pointed out that their solution for the stellar halo density distribution
is not unique, thus the resolution of this puzzle may be in
multi-component models, such as those discussed below. 

Finally, we point out that although the model advocated by Bond10 assumes
no halo rotation, their data could not rule out net rotation at the level
of up to $\sim20$ km s$^{-1}$. The key systematic errors limiting the
precision are the distance scale errors, uncertain correction to the local
standard of rest, and systematic errors in radial-velocity and
proper-motion measurements (see their Section 5.3). Nevertheless,
\cite{Smith2009MNRAS} did not detect halo rotation from their sample with
more robust proper-motion measurements; similarly
\cite{Allende06} found no evidence for halo rotation using SDSS radial velocities.

\subsection{Beyond a Simple Power Law: One Halo, Two Halos, Many Halos?   \label{Sec:OuterHalo}}

\begin{figure}%
\centerline{\psfig{figure=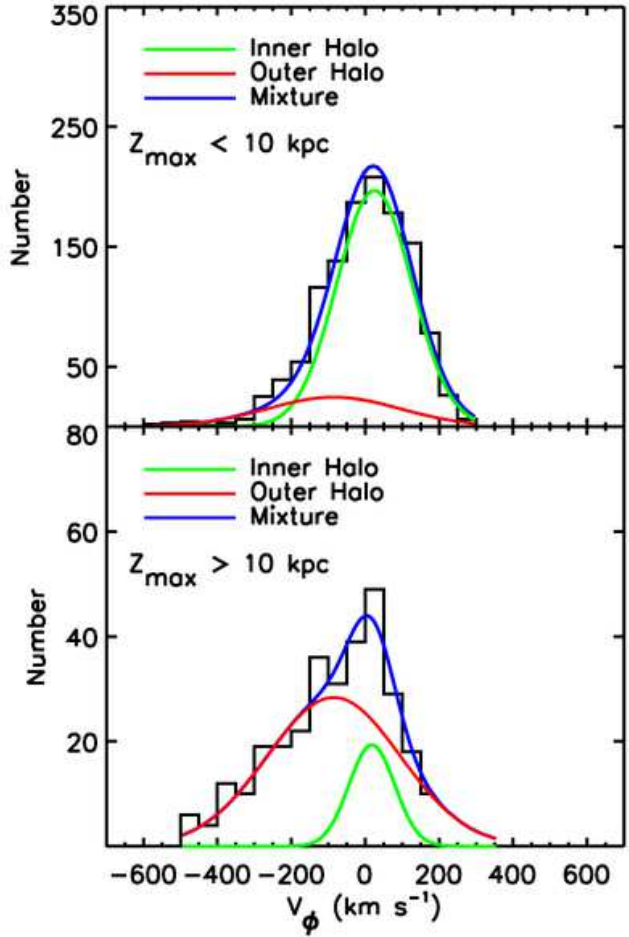,height=35pc}}
\vskip -0.8in 
\caption{
Rotational properties for the low-metallicity ($[Fe/H]< -2.0$) subsample of
SDSS calibration stars,
divided into stars with maximum orbital distances from the plane ($Z_{max}$) above 
or below 10 kpc (Figure 10 from \citealt{Carollo2010}).  The histograms show the observed 
distribution of rotational velocity ($v_\Phi$), and the smooth curves show models for
the inner (green) and outer (red) halo components (the model sum is shown by the blue
curves).}
\label{Cfig10}
\end{figure}

Due to the SDSS faint flux limit, the Juric08 results for the spatial distribution of
main-sequence halo stars are limited to the volume within $R\sim20$ kpc and
$|Z|\sim10$ kpc. Additional data suggest that the Juric08 single power-law halo
model cannot be extrapolated beyond these limits. First, a kinematic analysis of
halo stars, {within the same distance limits} by \cite{Carollo07,
Carollo2010}, suggests that the halo consists of two broadly overlapping structural
components, an ``inner halo" and an ``outer halo" (see Figure~\ref{Cfig10}).  
These labels are not merely descriptors for the regions studied, but rather are 
labels for two individual stellar populations.  These components exhibit
different spatial density profiles, stellar orbits, and stellar metallicities,
with the inner halo to outer halo transition occurring, according to these authors,
at Galactocentric distances of 15-20 kpc. This result follows
from their kinematic analysis of SDSS calibration stars within 4 kpc (including stars other
than main-sequence dwarfs), and is not an {in situ} measurement.
The inner halo was shown to comprise a population of stars exhibiting a
flattened spatial density distribution, with an inferred axial ratio on the
order of $q_H\sim0.6$ and $n=3.2\pm0.2$, no rotation at the level of $\sim$10 km
s$^{-1}$, and a metallicity distribution peaked at $[Fe/H]\sim-1.6$. These
properties of the inner halo are in very good agreement with the results from
Juric08, Ivezic08, and Bond10, based on direct mapping (as opposed to indirect inferences in
the Carollo et al. analysis). The outer halo comprises stars that exhibit a more
spherical spatial density distribution, with an axial ratio $q_H\sim0.9$ and
$n=1.8\pm0.3$, a clear retrograde net rotation ($\langle v_{\Phi}\rangle$ 
$\sim -80$ km s$^{-1}$), and a metallicity distribution peaked at $[Fe/H]\sim-2.2$.  

The Carollo et al. results were recently questioned by \cite{Schoenrich2011},
who argued that distance errors resulting from luminosity biases,
and/or improper accounting for measurement errors and the use of
Gaussian fitting resulted in a distorted identification of the halo components. 
They re-evaluated the same data and failed to detect ``any reliable evidence for a
counter-rotating halo component.'' In a rebuttal of these claims, 
\cite{BeersRebuttal} re-analyzed their original dataset (re-classifying the 
main-sequence turnoff stars that were the primary source of concern for
Sch\"onrich et al.), pointed out that Sch\"onrich et al. had themselves adopted
an incorrect main-sequence luminosity relationship from Ivezic08 (which significantly
affected their interpretation), and confirmed the presence of a lower
metallicity counter-rotating halo (whether or not the turnoff stars in question
were used in the analysis). The Beers et al. paper also provided additional
evidence for the presence of an inner/outer halo dichotomy, based on other
datasets and methods of analysis. They concluded that ``Ultimately, geometric
distances from Gaia for stars in the halo populations will eliminate any
remaining questions concerning the impact of uncertain photometric parallaxes on
these conclusions. However, our view is that presently available data already
reject the single-halo interpretation beyond reasonable doubt.''

Ideally, to resolve these ambiguities the spatial distribution of halo stars,
and their kinematic and chemical behavior, should be measured { in situ}
using main-sequence stars. Unfortunately, even turnoff stars with $M_r=5$ would
be as faint as $r=25$ at a distance of 100 kpc. Until the advent of LSST and
Gaia (for robust calibration of the photometric parallax relation) surveys (see
\S\ref{roadahead}), it will not be possible to perform such measurements over a
large area of sky. Nevertheless, several studies over small sky areas, or using
tracers more luminous (but less numerous!) than main-sequence stars, have
provided further, and often intriguing, insights into the properties of 
the halo. 

The analysis of de Jong et al. (2010), based on a color-magnitude diagram
fitting approach using templates of old stellar populations with differing
metallicities, produced a sparse three-dimensional map of the stellar
distribution of SDSS main-sequence turnoff stars within $r \sim 30$ kpc, derived
from the ten ``vertical'' (in Galactic coordinates) photometric scans of width
2.5$^\circ$ obtained during the SEGUE sub-survey of SDSS-II. Their Figure 6
provides clear {in situ} evidence for a shift in the mean metallicity of the
Milky Way's stellar halo -- within $r \sim 15$ kpc their derived stellar
halo exhibits a mean metallicity of $\langle [Fe/H]\rangle \sim -1.6$, changing
to $\langle [Fe/H]\rangle \sim -2.2$ at larger Galactocentric distances. In
addition, inspection of the spatial density profiles of their template
populations (their Figure 7) suggests rather different spatial behaviors for
their ``inner-halo like'' template population and that of their ``outer-halo
like'' template population. Their derived inner-halo density profile falls off
rapidly with distance from the Galactic center to $r \sim 15-20$ kpc; beyond
this region a substantially lower density, slowly varying, outer-halo density
profile was found. Note that the de Jong et al. analysis was restricted to
distances $r < 30$ kpc. When a single power-law was fit to this entire region
they obtained an index of $n = 2.75 \pm 0.07$, in excellent agreement with the
previous work of Bell et al. (2008) and Juric08.
 
Over most of the sky observed by SDSS, the distribution of blue
main-sequence stars can be mapped out to a distance limit of $\sim$20-30
kpc. However, in about $\sim$300 deg$^2$ of sky from the so-called Stripe
82 area, co-added imaging based on multiple observations has a limiting
magnitude about two mags fainter ($r\sim24$) than single-epoch SDSS data,
and can be used to map the number-density distribution of blue
main-sequence stars out to $\sim$40 kpc. \cite{Sesar2010} analyzed this
dataset and found that it agrees well with the extrapolations of the
Juric08 model to Galactocentric distances less than $\sim$25 kpc. However,
at larger distances the model overpredicts the observed counts by about a
factor of two, strongly suggesting that the halo stellar number-density
profile becomes much steeper. Although Sesar et al. could not derive
precise quantitative adjustments to the Juric08 model parameters, they
approximately estimated a change of the halo power-law index from $n\sim3$
to about $n\sim5$. They also detected a decrease in the median metallicity
between Galactocentric radii of 10 kpc and 20 kpc of 0.02 dex kpc$^{-1}$, a
factor of four larger than the upper limit of 0.005 dex kpc$^{-1}$
determined by Ivezic08 within 10 kpc. Extrapolation of this gradient to a
Galactocentric radius of $\sim$50 kpc would result in a metallicity value
similar to that determined for the outer halo by Carollo et al. Note that,
in the interpretation of Carollo et al., this ``gradient" is actually the
result of the lessening degree of importance of the inner halo, relative to
the more metal-poor outer halo, as one moves outward. Once the outer-halo
population dominates, there is no expected decline in metallicity.

\cite{SJI2011} used the Canada-France-Hawaii Telescope Legacy Survey (CFHTLS)
data, covering 170 deg$^2$, to study the distribution of near-turnoff
main-sequence stars along four lines of sight to heliocentric distances of
$\sim$35 kpc. They found that the halo stellar number-density profile
becomes steeper at Galactocentric distances greater than $\sim$28 kpc, with the
power-law index changing from $n= 2.62\pm0.04$ to $n=3.8\pm0.1$. They
measured the oblateness of the halo to be $q_H=0.70\pm$0.01 (statistical
error only) and detected no evidence of it changing across the range of
probed distances, nor any changes in the median metallicity.

\cite{DBE2011} explored similar issues, using a sample of
$\sim$20,000 BHB and blue straggler stars detected by SDSS over 14,000
deg$^2$ of sky, and obtained almost identical results to those from
\cite{SJI2011}, based on main-sequence stars. Their best fitting model
has an inner power-law index of $n=2.3$ and an outer index of $n=4.6$, with
the transition occurring at $\sim$27 kpc, and a constant halo flattening of
$q_H=0.6$. They concluded that ``the stellar halo is composed of a smooth
underlying density, together with some additional substructures such as the
Virgo Overdensity and the Sagittarius Stream''. In addition, the
distribution of RR Lyrae stars from the SEKBO survey \citep{Keller2008},
and of RR Lyrae stars from SDSS Stripe 82 data \citep{Sesar2010,
Watkins2009}, indicates a steeper density profile beyond 30 kpc.
Furthermore, using the LONEOS sample of RR Lyrae stars, \cite{Miceli2008}
argued for the presence of a dual halo in order to account for the
apparently very different spatial profiles of Oosterhoff Type I and
Oosterhoff Type II subsamples. { Taken together, these studies provide
strong {in situ} support for rejecting the single-halo hypothesis.}

\subsection{What is the Nature of the Outer Halo?}

Despite the growing evidence that stellar distribution in the Milky Way halo is
more complex than a smooth single power law, and may well comprise at least two
primary stellar populations, quantitative knowledge about the most distant parts
of the Milky Way is still limited. The smooth analytic descriptions of the
stellar distribution discussed above begin to fail beyond 30 kpc from the
Galactic center due to the presence of rich substructure, as discussed in the
next section. Even when these deviations are ignored, only qualitative
statements can be made about the spatial and metallicity distributions of halo
stars -- beyond 30 kpc, the spatial distribution is probably steeper than a
$1/r^3$ power law, and the median metallicity is likely lower by 0.3-0.5
dex than the value of $[Fe/H] \sim -1.5$ representative for the inner halo
(based on BHB stars, see Xue et al. 2008; Beers et al. 2011). Indirect evidence
suggests that the outer halo possesses a retrograde net rotation; this
result should be checked by further study.  

More robust results exist for the behavior of the halo radial-velocity
dispersion out to $\sim$100 kpc. \cite{Battaglia05} used a heterogeneous
sample of 240 objects (globular clusters, satellite galaxies, BHB stars,
red giants), and measured a decrease in radial-velocity dispersion from 120
km s$^{-1}$ at 30 kpc to about 80 km s$^{-1}$ at 100 kpc (these authors
also claim 50 km s$^{-1}$ at 120 kpc, but there are only four objects
beyond 100 kpc in their sample). Using about 2400 BHB stars detected by
SDSS out to 60 kpc, \cite{Xue08} measured a slightly lower velocity
dispersion than Battaglia et al. (e.g., $\sim$100 km s$^{-1}$ at 30 kpc vs.
120 km s$^{-1}$), and a shallower gradient by about a factor of two (e.g.,
their best fit implies a drop of 18 km s$^{-1}$, or 18\%, between 30 kpc
and 100 kpc, compared to a drop of 40 km s$^{-1}$, or 33\%, in Battaglia et
al.). \cite{Brown2010} used a mix of 910 BHB and blue straggler stars from
the Hypervelocity Star survey to measure the halo radial-velocity
dispersion out to 75 kpc. They obtained results in statistical agreement
with \cite{Battaglia05} and \cite{Xue08}, which they summarized as (see
their Figure 7 for a pictorial summary) ``the Milky Way radial-velocity
dispersion drops from $\sigma=110$ km s$^{-1}$ at $R_{gc}=15$ kpc to
$\sigma=85$ km s$^{-1}$ at $R_{gc}=80$ kpc.'' ($R_{gc}$ is the
Galactocentric radius). 

In contrast, \cite{DePropris2010} used $\sim$700 BHB stars from the 2Qz
Redshift Survey, and found a very strong {increase} in the radial-velocity
dispersion from 100 km s$^{-1}$ at 30 kpc to 200 km s$^{-1}$ at 80-100 kpc.
Furthermore, their dispersion value of $\sim$150 km s$^{-1}$ at 60 kpc is
significantly different from the 94 km s$^{-1}$ obtained by \cite{Xue08}
using {the same tracer population}. The value of $\sim$150 km s$^{-1}$ at
60 kpc seems completely ruled out by other studies (e.g., see Figure 7 in
\citealt{Brown2010}). 

Currently, there are no {in situ} measurements of tangential velocity
dispersion for stars from the outer halo. As discussed above, the
difference between $\sigma_\Phi$ and $\sigma_\theta$ is an important
measurement for understanding the gravitational potential and stellar
distribution in the halo. It seems that such measurements will be possible
for post-main-sequence stars with Gaia. For example, a star with $M_r=1$
would have SDSS $r$-band magnitude of 19 at a distance of 40 kpc. The
expected proper-motion error from Gaia is $\sim$0.1 mas yr$^{-1}$ for such
a star (see \S\ref{roadahead}), and corresponds to a velocity error of 20
km s$^{-1}$. Such an error is sufficiently small to enable detailed mapping
of $\sigma_\Phi$ and $\sigma_\theta$, based on enormous numbers of distant
BHB and red-giant stars. 

The chemistry of the outer-halo population, and its differences with
respect to the inner-halo population, has only begun to be explored.
Present evidence indicates that the diversity of stellar abundance ratios
is far greater for members of the outer halo than for the inner halo (see the
Introduction of \citealt{Carollo2011}). The most striking chemical
differences between the inner- and outer-halo populations may be revealed
by the recently recognized contrast in the frequency of so-called
carbon-enhanced metal-poor (CEMP) stars by \cite{Carollo2011}. These
authors have argued that the previously recognized increase in the
frequency of CEMP stars with declining metallicity is due to the fact that
the outer-halo component of the Galaxy possesses about twice the fraction
of CEMP stars, relative to carbon-normal stars, at a given low metallicity,
than the inner-halo component. In their view, the observed
correlation is a manifestation of the lower metallicity of outer-halo
stars, which begin to dominate halo samples at low abundance. This idea can
also account for the observed increase in the fraction of CEMP stars, at a
given metallicity, as a function of height above the Galactic plane
\citep{Frebel2006, Carollo2011}, a result that would be difficult to
understand in the context of a single-halo population.

In summary, the outer parts of the halo, beyond $\sim30$ kpc from the
Galactic center, probably have a steeper density distribution ($n>3$) and
lower median metallicity ($[Fe/H]<-1.5$) that the inner halo. The outer
halo appears to be less ``squashed'' than the inner halo, likely exhibits a
net retrograde rotation, and its radial-velocity dispersion probably
decreases with Galactocentric distance. There appear to be clear
differences in the chemistry of outer-halo stars relative to those of the
inner halo. And, as discussed in the next section, both fine and coarse
substructure in the outer halo appears much more prominent than in the inner
halo.

\subsection{Streams and Other Substructure \label{Sec:substructure}}

\begin{figure}%
\vskip -2.6in 
\centerline{\psfig{figure=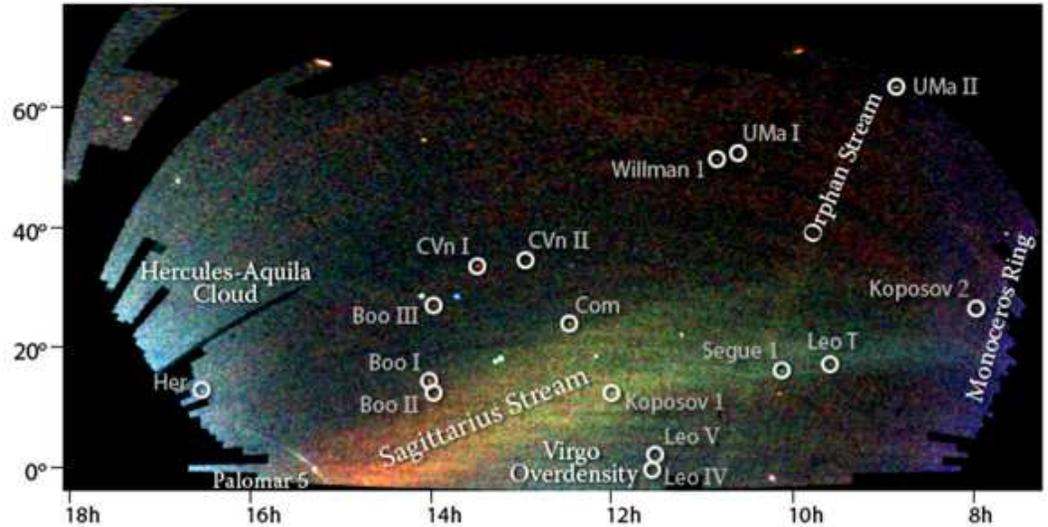,height=45pc}}
\vskip -2.1in 
\caption{
The ``Field of Streams'' map of blue stars in the outer regions of the Milky Way, derived
from the SDSS images of the northern sky, shown in a Mercator-like
projection of equatorial coordinates (based on Figure 1 from \citealt{Belo2006a};
courtesy of V. Belokurov, Institute of Astronomy, Cambridge).
The color indicates the distance of the stars (of the order 10 kpc), with red being the 
most distant and blue being the closest, while the intensity indicates the density of 
stars on the sky. There are several
structures visible in this map, as marked, that demonstrate the halo
is not a smooth structure. Circles enclose new Milky Way companions
discovered by the SDSS, as labeled; two of these are faint globular star clusters, 
while the others are faint dwarf galaxies.}
\label{FoStreams}
\end{figure}

Was the Milky Way halo, or at least its outer parts, actually assembled from
many merged satellite galaxies? Within the framework of hierarchical galaxy
formation \citep{Freeman2002}, the spheroidal component of the luminous
matter should reveal substructures, such as tidal tails and streams
\citep{jhb96,hw99,bkw01, har01}. Substructures are expected to be
ubiquitous in the outer halo (Galactocentric radii beyond 15-20 kpc), where
the dynamical timescales are sufficiently long for them to remain spatially
coherent \citep{jhb96,may02}, and indeed many have been discovered during
the last decade, the famous ``Field of Streams'' \citep{Belo2006a} is shown
in Figure~\ref{FoStreams}. 

The tidal streams of the disrupting Sagittarius dwarf spheroidal galaxy
\citep{Ibata1994} were the first ones discovered\footnote{We note that
no clear stellar component has been associated with the Magellanic Stream yet, 
despite searches for it \citep{MooreDavis1994,Westerlund1990,Cioni2011,Saha2010}.} \citep{ive00,yan00,
Vivas2001,new03}, and they are still the best examples of such substructures, with the streams wrapping
around most of the sky \citep{Majewski2003,Ibata2001}. Other known substructures
include the Virgo Stellar Stream \citep{Duffau2006,Prior2009}, and several other
small and large structures associated with the Virgo overdensity
\citep[][Juric08]{new07,Vivas2008}, 
the Monoceros structure \citep[][Ivezic08]{yan03,RochaPinto2003,Ibata2003},
the Triangulum-Andromeda overdensity \citep{RochaPinto2004},
the Hercules-Aquila cloud \citep{Belo2007b},
the Pisces overdensity \citep{Sesar2007,Sesar2010,SesarPisces,Watkins2009,Juna2009},
the Orphan stream \citep{Belo2007c}, 
and other smaller overdensities \citep{new02,Belo2007a,Belo2006b,CK2006,Stark2009,vz06}
and streams \citep{GD06,Grillmair09,Schlaufman09,Klement09,Klement2010}.
Similar abundant substructure has been detected in the M31 halo \citep{Ferg2002,IbataM31}. 
Last but not least, the SDSS imaging data have enabled a large number of new discoveries
of dwarf galaxy companions to the Milky Way, with luminosities as small as 10$^{-7}$ of
the Milky Way luminosity (for an up-to-date review, see \citealt{Willman2010}). 

\begin{figure}%
\vskip -0.0in 
\centerline{\psfig{figure=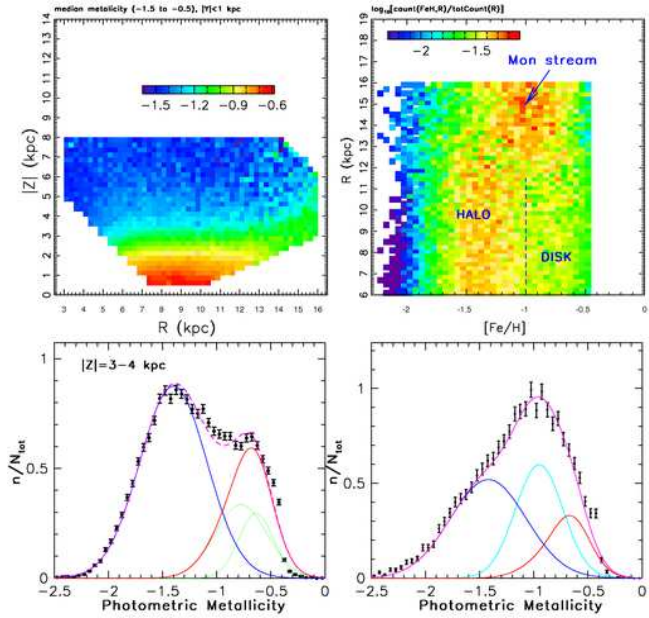,height=20pc}}
\vskip -0.1in 
\caption{Figure 18 from \cite{Ivezic08}. 
Top left panel: Dependence of the median photometric metallicity for
one million stars with $14.5 < r < 20$, $0.2<g-r<0.4$, and 
$|Y|<1$ kpc, in cylindrical Galactic coordinates $R$ and $|Z|$.  This
$Y$ range is selected to include the Monoceros stream, which
represents an overdensity by a factor of $\sim$1.5 in a region
around $R\sim15$ kpc and $|Z|\sim 3-4$ kpc. As discernible from the
map, this region has a larger median metallicity than expected for
this $|Z|$ range based on extrapolation from smaller R. Top right
panel: Conditional metallicity probability distribution for a
subsample of  $\sim$111,000 stars with $3 < |Z|/{\rm kpc} < 4$. The 
strong overdensity at $R > 12$ kpc is the Monoceros stream. The bottom 
panels show the metallicity distribution (symbols with error bars) for
a subsample of $\sim$40,000 stars with $6 < R/{\rm kpc} < 9$ (left)
and for $\sim$12,000 stars with $13 < R/{\rm kpc} < 16$ (right). The 
lines represent empirical fits based on Gaussian decomposition from \cite{Ivezic08}
(blue lines for halo component and red lines for disk component).   
The cyan line in the bottom right panel is a 0.22 dex wide Gaussian centered on
$[Fe/H]=-0.95$. It accounts for 33\% of stars in the sample that
presumably belong to the Monoceros stream.
}
\label{TIIf18}
\end{figure}

These substructures exhibit various degrees of contrast with respect to the
background counts, metallicity, and kinematic distributions of the smooth
halo. \cite{Bell2010} used a novel method for investigating differences in
age and metallicity between different halo substructures: the ratios of
counts of BHB stars to main-sequence turnoff stars. Using SDSS data across
a quarter of the sky, they found large variations of this ratio, with some
halo features almost completely devoid of BHB stars. The Monoceros stream
is a good example of substructure that simultaneously deviates from the
background distributions in all three observables (see Figure~\ref{TIIf18}). 
Originally interpreted as a ring around the Galaxy \citep[][]{yan03,
Ibata2003}, it was later revealed as an inclined stream that rotates faster
than the surrounding stars, and has a metallicity distribution between
those of disk and halo stars \citep[][Juric08, Ivezic08]{RochaPinto2003}. 

The discoveries of abundant substructure represent ample evidence that the
Milky Way halo is a very complex structure that holds important clues for
deciphering the processes that governed its formation and evolution. Yet,
despite this menagerie of substructure, we still do not have a consensus
answer for as simple a question as ``What fraction of halo stars was
accreted from merged galaxies?'' Of course, questions about the luminosity
and metallicity distributions of the merged galaxies are even more open. A
part of the problem is that, within the neighborhood of strong individual
substructures, such as those listed above, it is hard to define the
``smooth'' background, and when strong substructure is absent, it is hard
to find weaker features. Several approaches to quantifying the amount of
halo substructure have been taken in recent years. 

\begin{figure}%
\vskip -0.5in
\centerline{\psfig{figure=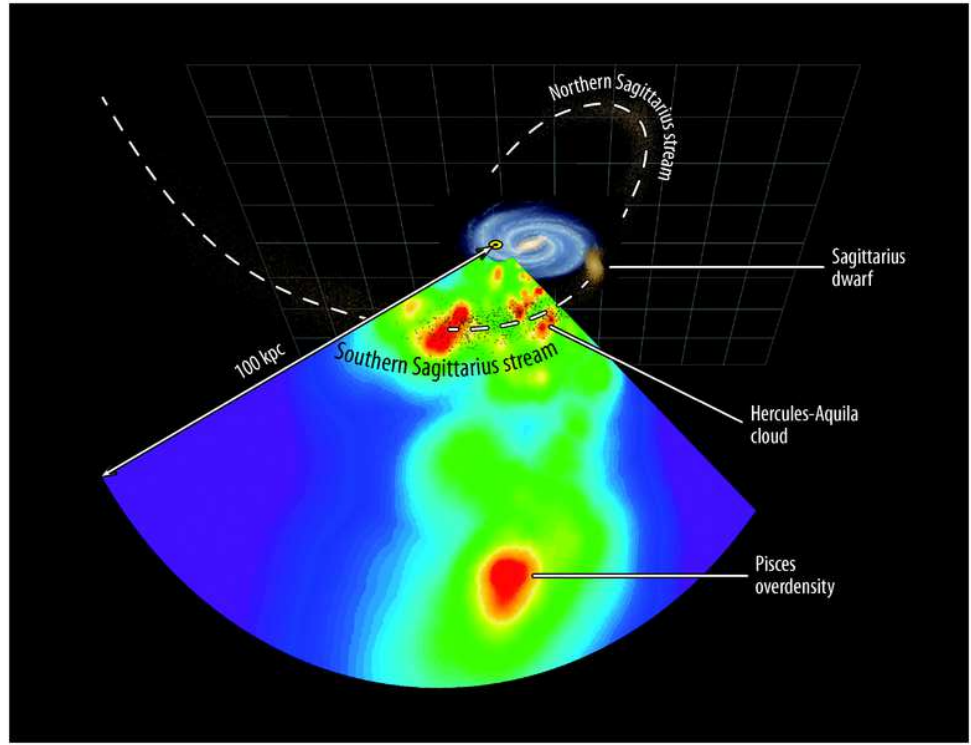,height=24pc,angle=90}}
\vskip -0.3in
\caption{The distribution of RR Lyrae stars from SDSS Stripe 82, contrasted with 
an artist's concept of the disk plane (based on Figure 12 from \citealt{Sesar2010}). 
The color scheme displays the RR Lyrae number density multiplied 
by the cube of the Galactocentric radius (logarithmic scale, from light blue to red). Note the rich
structure present. The white dots, outlined by white dashed lines, show the Sagittarius 
dSph (``Sgr dwarf'') and its tidal streams, as modeled by the \cite{Law2005} ``spherical'' 
model (the model stream overlaps with one of the detected clumps, ``Sgr for discussion see 
\citealt{Sesar2010}). An animated version is available from the ARA\&A website 
(courtesy of A. Mejia and B. Sesar).}
\label{SesarFig12}
\end{figure}

Three studies compared the spatial distribution of SDSS stars, with
distances based on photometric parallax methods, to estimates of the smooth
background, and used variations (the root-mean-square deviation, rms
scatter) of the measured density around that background as a quantitative
measure of substructure. The Juric08 approach used their best-fit smooth
models for the background estimate, and ruled out significant clumpiness on
spatial scales comparable to the pixel size of their maps (ranging from 25
pc to 500 pc). On the other hand, \cite{Bell2008} compared the spatial
distribution of SDSS stars to that from simulations where the halo is
composed {entirely} of disrupted satellites, and found them to be similar
for Galactocentric radii less than 40 kpc. They argued that no smooth model
can describe the data, and concluded that the stellar halo is dominated by
substructure, with the rms scatter, relative to smooth models on spatial
scales above 100 pc, of at least 40\%. The Bell et al. study relied on SDSS
observations of main-sequence stars and pushed the data all the way to its
faint limit ($r\sim22.5$). \cite{DBE2011} instead used BHB and
blue-straggler stars from SDSS DR8 to study their distribution to a similar
distance limit of 40 kpc, but with much brighter stars. They
reported for their smooth models that the rms scatter of the data around
the maximum likelihood model typically ranges between 5\% and 20\%. They
concluded that ``This indicates that the Milky Way stellar halo, or at
least the component traced by the A-type stars in the SDSS DR8, is smooth
and not dominated by unrelaxed substructure.''.  It should be kept in mind
that this inference reflects the present situation, and may not apply to
the situation at earlier times.  Using RR Lyrae stars from
SDSS Stripe 82, \cite{Sesar2010} found that ``At least 20\% of halo stars
within 30 kpc from the Galactic center can be statistically associated with
substructure.'' And further, that ``...beyond a Galactocentric distance of $\sim$30
kpc, a larger fraction of the stars are associated with substructure.''
(see Figure~\ref{SesarFig12}). 

The addition of kinematic data increases the contrast ratio relative to the
smooth background when searching for substructure. Using data from the
SEGUE spectroscopic survey, \cite{Schlaufman09} have shown that metal-poor
main-sequence turnoff stars within $\sim$20 kpc from the Sun exhibit clear
evidence for radial velocity clustering on very small spatial scales
(dubbed ``ECHOS'' for Elements of Cold HalO Substructure; see the bottom
panel in Figure~\ref{TIIIf21}). They estimated that about 10\% of the inner
halo turnoff stars belong to ECHOS, and inferred the existence of about
1000 ECHOS in the entire inner halo. Their ``result suggests that the level
of merger activity has been roughly constant over the past few Gyr and that
there has been no accretion of single stellar systems more massive than a
few percent of a Milky Way mass in that interval.'' \cite{Schlaufman2011}
argue that the most likely progenitors of ECHOS are dwarf spheroidal
galaxies. Typical values of metallicity ($[Fe/H] \sim -1$) and
radial-velocity dispersion ($\sim$20 km s$^{-1}$) for ECHOS imply a dwarf
galaxy mass of about 10$^9$ M$_\odot$. Theoretical predictions that
prominent halo substructures are likely to be metal-rich
\citep{Bullock2005, Font2008} are consistent with the typical ECHOS
metallicity, as well as with the measurements reported for the Monoceros
stream ($[Fe/H]=-1.0$, Ivezic08) and the trailing part of the Sagittarius
tidal stream ($[Fe/H]=-1.2$, \citealt{Sesar2010}). 

\cite{Smith2009MNRAS} used the full phase-space coordinates for a sample
of some 1700 SDSS subdwarfs, and found evidence for four discrete
overdensities localized in angular momentum space, which they dubbed Sloan
Kinematic Overdensities (SKOs). One of them was identified earlier in the
pioneering work by \cite{Helmi1999}, and two new substructures contain
stars that are localized in both kinematics and metallicity. One of them
has metallicity lower than that of halo background by $\sim$0.5 dex, and
appears to be related with an association of four globular clusters
(NGC5466, NGC6934, NGC7089/M2 and NGC6205/M13), suggesting that they may
have been part of the same accretion event. If so, then this implies that
the progenitor must have been a large satellite, similar in size to Fornax. 

\cite{Xue2011} analyzed kinematic data for a sample of over 4000 BHB stars 
detected by SDSS at distances 5-40 kpc. Using a method developed for the
analysis of data from the Spaghetti project \citep{Stark2009}, they found
an excess of stars that are both close neighbors and have similar radial
velocities, compared to a distribution expected for a random sample.
Notably, the excess is larger for a subsample of stars at distances beyond
20 kpc than for the closer subsample. Analogous analysis of mock catalogs
from simulations in which the stellar halo is composed entirely of
disrupted satellite debris exhibits a similar, though somewhat less prominent,
level of structure. In a separate study, also based on SDSS observations of
BHB stars and using a similar analysis method, \cite{Cooper2011} analyzed a
large number of state-of-the-art models for the stellar halo. They also
found that, for the inner halo, the models predict stronger clustering than
observed, suggesting the existence of a smooth component not currently
included in their simulations. 

In summary, these new studies consistently reveal that the inner region of
the Milky Way's stellar halo, within 30 kpc or so, definitely exhibits
substructure. Estimates of the fraction of stars that belong to
substructures cluster around a few tens of percent. Observations of several
luminous tracers of the outer region of the halo, such as RR Lyrae and BHB
stars, suggest that the substructure becomes more prominent with increasing
distance from the Galactic center, and that the fraction of stars belonging to
substructures is higher than for the inner halo. The available data (still)
cannot reliably exclude the possibility that practically {all} stars in
the outer halo belong to substructures. 

\section{UNANSWERED QUESTIONS}

It is not difficult to appreciate the progress that has been made, compared
to the state of the field and the open questions of only a decade ago. The
metallicity distribution of the halo was found to possess a very
low-metallicity tail, one that is correlated with kinematics and points to
an existence of a separate outer-halo component. The density profile of the
halo was shown to be more complex than originally thought, becoming steeper
beyond $\sim 25$kpc, and exhibiting significant substructure. Tens of dwarf
galaxies and a number of new streams, some with clear dwarf galaxy
progenitors, have been found, proving that the Sgr stream, while still
remaining the largest, was in no way a qualitatively unique event. And
finally, it has become clear that the thin and thick disks of the Milky Way
are demonstrably distinct physical components, separable by their
kinematics and $[\alpha/Fe]$ ratios, and mapped and measured in exquisite
detail. Thus we argue that the past decade of large surveys
has successfully retired issues \#1 through \#4, as well as issue \#6, mentioned
in Section~\ref{decadeAgo}.

However, as usually happens, these breakthroughs have left us with new
puzzles and questions to ponder over the decade to come. Here, we call
attention to only a few:
\begin{enumerate}
\item What is the nature, and the formation mechanism, of the two chemically
and kinematically separate disk populations?  On the most basic level,
is the thick disk a result of one or more merger events, or is it a 
natural consequence of secular evolution and radial migration? If it is the
latter, what explains the counter-rotating disks seen in some external galaxies?
\item If the thick disk was formed by mergers, how massive and numerous were
they?  What fraction of the thick disk material (if any) has been accreted,
and what fraction came from heating of the material already settled into the
thin disk (both gas and stars)?  Is there even a single thick disk to speak of,
or are there multiple intertwined populations tracing their origin to individual merger
events?  Could this be the explanation of the metal-weak thick disk and the
non-Gaussianity of $[Fe/H]$ and $[\alpha/Fe]$ distributions?
\item How do the properties of the bulge compare to those of the disk and
the halo? 
%
\item Will the inner/outer halo dichotomy be confirmed by {in-situ}
measurements?  Currently, our inferences about the outer-halo population are
largely drawn from local kinematically-selected samples. {In-situ} measurements of
the properties of halo stars at $r>25$ kpc, using large and representative
samples, may settle these controversies.
\item Assuming the outer halo is distinct from the inner halo, what are their
origins and mechanisms of formation? What (if any) fraction of the outer halo
has formed stars {in-situ}, as opposed to having accreted them? What
fraction of the inner halo traces its origin to merger events? What does
this tell us about the merger history of the Galaxy?
\item And finally, what is the gravitational potential of the Milky Way's dark matter
halo?  Multiple lines of evidence currently point to its near-sphericity, a result
at odds with expectations for a typical dark matter halo (either prolate or
oblate, with $q \sim 0.6$) from N-body simulations.
\end{enumerate}

\section{THE ROAD AHEAD  \label{roadahead}}

The last decade has seen fascinating observational progress in Milky Way studies. Nevertheless, the 
results discussed here will be greatly extended by several upcoming 
large-scale, ground-based projects, including APOGEE, LAMOST, the SkyMapper, Dark Energy Survey, 
Pan-STARRS, and ultimately the Large Synoptic Survey Telescope. These new 
surveys\footnote{Regretfully, due to space constraints we do not address
infrared surveys, such as WISE \citep{WISE2010}, GLIMPSE
\citep{GLIMPSE2009}, VISTA \citep{VISTA2006}, and UKIDSS \citep{UKIDSS2007}.
These surveys will have a major impact on studies of the Galactic bulge and
the dust-obscured regions of the Galactic plane.} will extend the faint
limit of the current surveys, such as SDSS, by up to 5 magnitudes. In
addition, the upcoming Gaia space mission will provide superb astrometric and
photometric measurement accuracy for sources with $r<20$, and will enable
unprecedented science programs. We briefly describe these new surveys, and
some of the impact they are expected to have on Milky Way studies. 


\subsection{SDSS APOGEE Survey \label{Sec:APOGEE}}


The SDSS-III Apache Point Observatory Galaxy Evolution Experiment (APOGEE)
will soon yield unprecedented insight into the chemical and kinematic
properties of the main Galactic components, with a unique dataset
for studying the bulge \citep{SDSSIII,SM2010,APOGEE2008,
Rockosi2009astro2010}. The APOGEE project is a three-year high-resolution
near-infrared spectroscopic survey that will target over 200 field centers
covering about 1200 deg$^2$ of sky. The project utilizes a new
300-fiber-fed $H$-band (1.51 to 1.68 $\mic$) spectrograph with a resolution
of $R\sim20,000$, and expected signal-to-noise ratio of 100 per resolution
element, for stars with $H\sim12$ (with the ARC 2.5m telescope used by SDSS).
The goal of the survey is to derive precision radial velocities ($\sigma<1$
km s$^{-1}$) and abundances ($\sigma<0.1$ dex) for about 100,000 stars,
targeted using the 2MASS imaging survey (red giants will be observable all
the way to the Galactic center). Abundances of 15 different elements, including
$Fe$, $C$, $N$, $O$, $\alpha$-elements, odd-$Z$ elements, and iron-peak
elements, will be measured. 

APOGEE will be the first spectroscopic survey to pierce through the dust
obscuration in the Galactic plane (extinction in the near-infrared $H$ band
is about six times smaller than in the optical $V$ band), and provide a
large, uniform database of chemical abundances and radial velocities for
stars across all of the known Galactic components. These data will provide
a robust set of constraints against which chemodynamical models for the
formation and evolution of the Galaxy can be tested. In particular, APOGEE
will provide powerful new constraints on the nature and influence of the
Galactic bar and spiral arms, and will conduct a legacy survey of Galactic
open clusters to constrain the history of star formation and chemical
enrichment of the Galactic disk
\citep{Frinchaboy2010}.

\subsection{The LAMOST Galactic Surveys  \label{Sec:LAMOST}} 

The Large Area Multi-Object fiber Spectroscopic Telescope (LAMOST) is a
4m-class telescope with 4000 optical fibers in the focal plane, sited at the
Xinglong Observatory in northeast China. This telescope, built by the
National Astronomical Observatory of China, will carry out a spectroscopic
survey of millions of Galactic stars over a five or six year period,
expected to start in 2012. The magnitude limit, wavelength range, and
spectral resolution of the LAMOST Galactic structure surveys will be
similar to that achieved by SEGUE/SDSS. 

The stellar science goals, grouped under the LAMOST Experiment for Galactic
Understanding and Exploration (LEGUE) effort, are divided into three major
parts: (1) The spheroid survey, (2) The Galactic anticenter survey, and (3)
The Galactic disk and open clusters survey. The stellar surveys will
receive about half of the available time, resulting in a sample of about 6
million bright disk stars, and at least 2 million fainter halo stars. These
surveys will result in the largest homogeneous spectroscopic datasets for
stars in the Milky Way, with several times more spectra than obtained
by the SDSS and RAVE surveys combined.

\subsection{SkyMapper, Pan-STARRS, and the Dark Energy Survey}

There are three imminent optical surveys that will cover large swaths of
the optical sky to faint limits, and are destined to yield many significant
discoveries. They are, in many ways, similar to the SDSS imaging survey
(including the photometric systems), but they will extend it significantly
in sky coverage, imaging depth, and temporal coverage. 

\subsubsection{SkyMapper}

The SkyMapper \citep{Murphy2009} is a 1.35m telescope with a 5.7 deg$^2$
field of view and a 0.27 Gigapixel camera. Its primary goal will be to
undertake the Southern Sky Survey: a six band, six-epoch (in each band)
digital record of the entire southern sky. The survey will provide
astrometry and photometry for objects with $8 < r < 23$. Each of the six
epochs will use 110 s exposures that will be about 1 mag shallower than
SDSS data, but when co-added, will reach the SDSS depth. The four red
bandpasses ($griz$) are designed to be similar to the SDSS bandpasses.
SkyMapper has two additional, distinctive ultraviolet filters, a
Str\"{o}mgren system-like $u$-band filter, and a unique narrow $v$-band
filter near 4000 \AA. These two filters bracket the Balmer jump in stellar
spectra, and are designed to efficiently identify metal-poor stars
\citep{Bessell2011}. The advertised performance requirements include a
photometric precision of 0.03 mag globally, and astrometric precision
(better than 50 mas) that will enable measured proper motions accurate to about 4
mas yr$^{-1}$ over the five year baseline of the survey. 

The SkyMapper's Southern Sky Survey will extend many of the results based
on SDSS imaging survey to the Southern hemisphere, over 20,000 deg$^2$ of
sky \citep{Keller2007}. The photometric parallax methods developed for SDSS
data should be directly applicable to SkyMapper's data, and photometric
metallicity methods should perform even better, thanks to the optimized
ultraviolet bandpasses. The expected proper-motion accuracy is essentially
the same as delivered by the SDSS-POSS proper-motion catalog, and thus it
will be possible to extend many of the SDSS-based studies described here to
essentially the entire sky.

\subsubsection{Pan-STARRS}

The Panoramic Survey Telescope And Rapid Response System (Pan-STARRS;
\citealt{Kaiser2010}) is a a wide-field, multi-filter, multi-epoch
astronomical survey program. The program is currently based on a 1.8m
telescope with a 7 deg$^2$ field of view and a 1.4 Gigapixel camera (PS1),
which began full science operations in 2010. The largest of the PS1 surveys
is the 3$\pi$ Survey, which is planned to cover the 30,000 deg$^2$ of sky
visible from Hawaii ($\delta>-30^\circ$) in five filters (SDSS-like $griz$
and $y$ at $\sim$1\mic), with pairs of observations in each filter being
taken at six different epochs. 

PS1 will increase the SDSS sky coverage by a factor of two, and will reach
at least a magnitude deeper (with a second telescope, or perhaps all four
that are envisioned in this program, the depth gain could be up to another
magnitude). Both of these advantages will most likely yield new discoveries
in the context of Milky Way studies. A additional magnitude of depth
corresponds to 60\% larger distance limit, and could bridge the 25 kpc to
40 kpc range, where the transition between the inner and outer halo is
probably taking place, with turn-off stars. In addition, the coverage of
the Galactic plane will be much better than with SDSS, and the $y$-band
will be more apt at penetrating through the high ISM dust extinction at low
Galactic latitudes. Unfortunately, the Pan-STARRS system does not include
an ultraviolet band required for photometric metallicity estimates.

\subsubsection{The Dark Energy Survey}

The Dark Energy Survey (DES) will utilize a new 0.52 Gigapixel camera at
the 4m Blanco telescope (3.8 deg$^2$ field of view) to cover 5000 square
degrees of the southern sky \citep{Flaugher08}. The survey will
be completed during a 5-year period starting in 2012, and will include
SDSS-like $griz$ passbands and the $y$-band. Similarly to Pan-STARRS, DES
will not include an ultraviolet band. As its name implies, although Milky Way
studies are not its primary goal, it will nevertheless provide valuable
data. 

Although the Dark Energy Survey will cover ``only'' about 5,000 deg$^2$ of
the Southern sky, it will reach about 1.5-2 magnitudes deeper than the
SkyMapper survey (and SDSS). This depth gain, and corresponding improvement
in the limiting distance by a factor of 2-2.5, is likely to bring
significant new discoveries, especially in the context of the Galactic halo.

\subsection{Gaia}

Gaia is an ESA Cornerstone mission set for launch in 2013. Building on
experience from HIPPARCOS, it will survey the sky to a magnitude limit of
$r\sim20$ (approximately, see below), and obtain astrometric and three-band
photometric measurements for about 1 billion sources, as well as
radial-velocity and chemical-composition measurements (using the 847-874 nm
wavelength range) for 150 million stars with $r<18$ \citep{Wilkinson05,
PerrymanGaia}. The final data product, the Gaia Catalogue, is expected to
be published by 2020, although early data releases are planned. 

Gaia's payload will include two telescopes sharing a common focal
plane, with two $1.7^\circ \times 0.6^\circ$ viewing fields separated by a
highly stable angle of $106.5^\circ$. The focal plane includes a mosaic of
106 CCDs, with a total pixel count close to one billion. Due to
the spacecrafts' rotation and precession, the entire sky will be scanned in TDI
mode (time-delay-and-integrate, or drift scanning) about 70 times, on
average, during 5 years of operations. Gaia will produce broad-band $G$
magnitudes with sensitivity in the wavelength range 330-1020 nm (FWHM
points at $\sim$400 nm and $\sim$850 nm). The spectral energy distribution
of each source will be sampled by a spectrophotometric instrument providing
low-resolution spectra in the blue ($BP$, effective wavelength $\sim$520
nm) and in the red ($RP$, effective wavelength $\sim$800 nm). In addition, the
RVS instrument (radial velocity spectrograph) will disperse the light in
the range 847--874 nm, for which it will include a dedicated filter. 

\subsection{LSST}

The Large Synoptic Survey Telescope (LSST) is the most ambitious currently
planned wide-field ground-based optical system \citep{Ivezic08LSST}. The
current baseline design, with an 8.4m primary mirror, a 9.6 deg$^2$ field
of view, and a 3.2 Gigapixel camera, will allow about 20,000 square degrees
of sky visible from Cerro Pach\'{o}n in Northern Chile to be covered to a
depth of $r\sim27.5$ over the 10-year survey. About 1000 observations
(summed over the six bands, $ugrizy$) will be obtained during that period,
enabling unprecedented time-domain studies. LSST will obtain proper-motion
measurements of comparable accuracy to those of Gaia at its faint limit
(and $r\sim$20), and smoothly extend the error vs. magnitude curve deeper
by about $5$~mag (see Figure~\ref{LSSTvsGaia}).  

LSST will produce a massive and exquisitely accurate photometric and
astrometric dataset for about 10 billion Milky Way stars. The coverage of
the Galactic plane will yield data for numerous star-forming regions, and
the $y$-band data will penetrate through the interstellar dust layer. With
its $u$-band data, LSST will enable studies of metallicity and kinematics
using the {same sample} of stars out to a distance of $\sim40$~kpc ($\sim
200$~million F/G main-sequence stars brighter than $r=23$; for a discussion
see Ivezic08), and the spatial distribution of halo turn-off stars will be
traced out to $\sim$100 kpc. No other existing or planned survey will
provide such a massive and powerful dataset to study the outer halo. The
LSST, in its standard surveying mode, will be able to efficiently detect RR
Lyrae stars, and hence explore the extent and structure of the halo out to 400
kpc (see Figure~\ref{LSSTvsSDSS}). All together, the LSST will enable
studies of the stellar distribution beyond the presumed edge of the
Galactic halo, of their metallicity distribution throughout most of the
halo, and of their kinematics beyond the thick disk/halo boundary
\citep[for more detailed discussion see the LSST Science Book;
][]{LSSTscibook}. 

\subsection{The Synergy between Gaia and LSST}

\begin{figure}%
\centerline{\psfig{figure=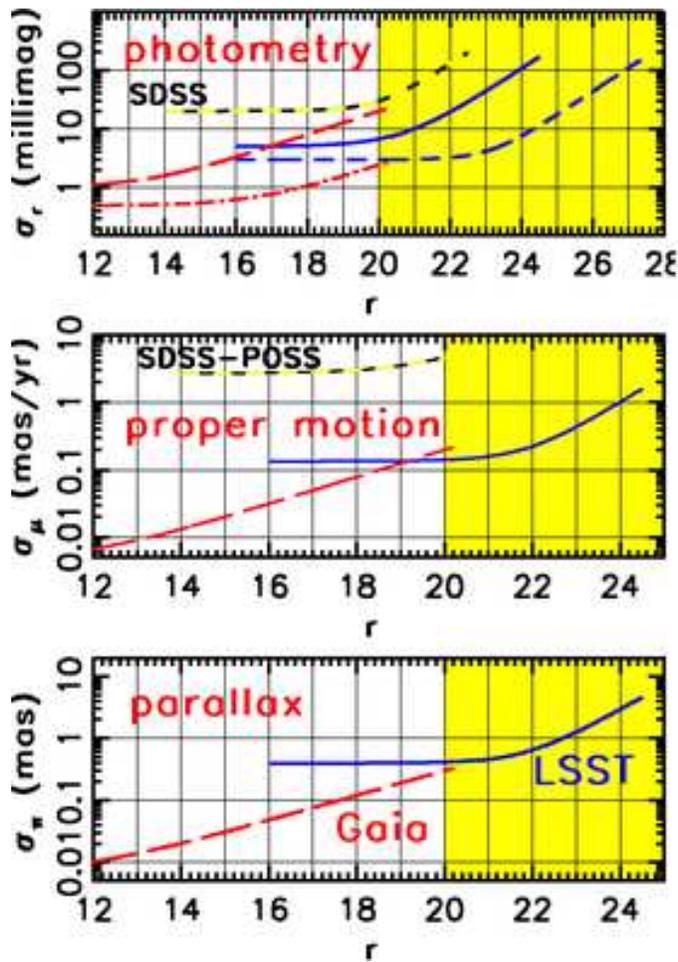,height=30pc}}
\vskip -0.2in
\caption{A comparison of the photometric, proper-motion, and parallax errors
for SDSS, Gaia, and LSST, as a function of apparent magnitude $r$, for  
a G2V star (Eyer et al, in prep). In the top panel, the curve marked ``SDSS'' corresponds
to a single SDSS observation. The red curves correspond to Gaia; the
long-dashed curve shows a single-transit accuracy, while the    
dot-dashed curve shows the end-of-mission accuracy (assuming 70 transits).
The blue curves correspond to LSST; the solid curve shows a single -visit accuracy,
while the short-dashed curve shows the accuracy for 
co-added data (assuming 230 visits in the $r$-band). The curve marked
``SDSS-POSS'' in the middle panel shows the accuracy delivered by the
proper-motion catalog of \citet{Munn04}.}
\label{LSSTvsGaia}
\end{figure}

\begin{figure}%
\centerline{\psfig{figure=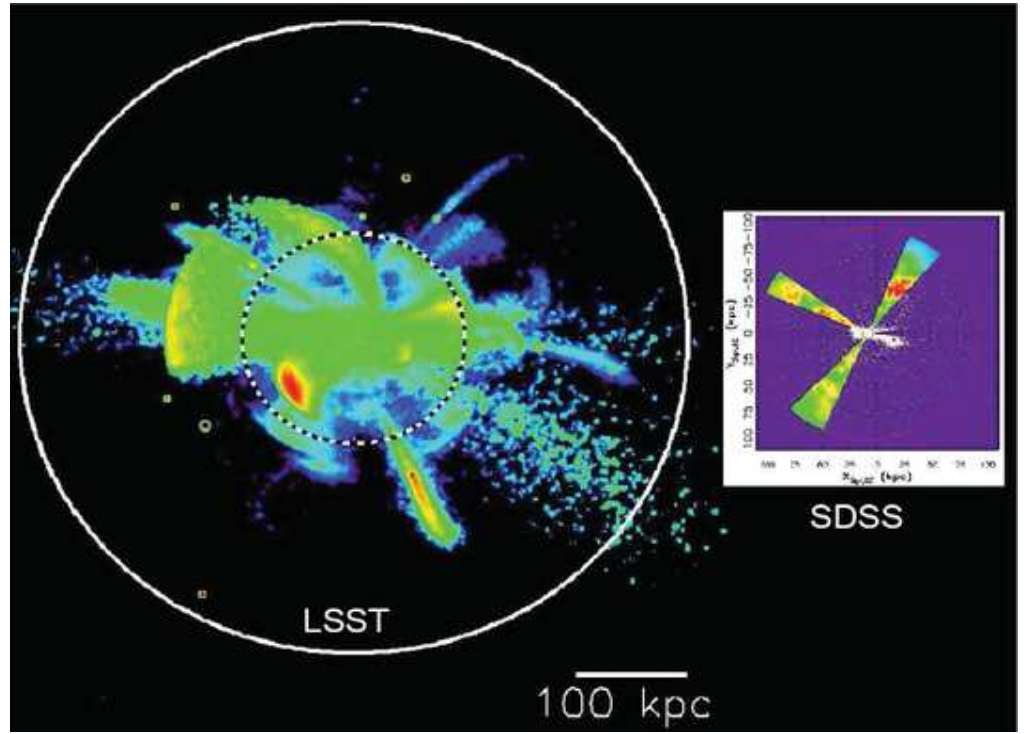,height=23pc}}
\vskip -0.2in
\caption{A simulation of the outer regions of the Milky Way (left), compared to the 
current state-of-the-art data (right), shown on the same spatial scale. The
data panel presents the number density multiplied by the cube of the
Galactocentric radius (logarithmic scale with dynamic range of 1000, from
blue to red), for $\sim$1000 SDSS RR Lyrae stars within 10$^\circ$ of the
Sgr dwarf tidal stream plane \citep{Ivezic2004}. The same color coding was
used to visualize the stellar number density for a Milky Way-type galaxy
simulation from \cite{Bullock2005}, shown on the left. Set within an
$\Lambda$CDM merger history, these simulations track the accretion and
disruption of hundreds of dwarf galaxies into Milky-Way size halos. With
LSST, RR Lyrae stars will be found beyond the presumed Milky Way tidal
radius ($\sim$300 kpc, white circle), and the much more numerous.
main-sequence stars will trace the structure out to 100 kpc (dashed circle)
The latter distance range can at present only be probed using RR Lyrae stars and
other rare non-main-sequence stars.}
\label{LSSTvsSDSS}
\end{figure}

In the context of Gaia, the LSST can be thought of as its deep complement.
A detailed comparison of LSST and Gaia performance is given in
Figure~\ref{LSSTvsGaia}. Gaia will provide an all-sky catalog with
unsurpassed trigonometric parallax, proper-motion, and photometric
measurements to $r\sim20$, for about $10^9$ stars. LSST will extend this map
to $r\sim27$ over half of the sky, detecting about $10^{10}$ stars. Because
of Gaia's superb astrometric and photometric quality, and LSST's
significantly deeper reach, the two surveys are highly complementary -- Gaia
will map the Milky Way's disk with unprecedented detail, and LSST will
extend this map all the way to the edge of the known halo and beyond.

\begin{figure}%
\vskip -0.2in
\centerline{\psfig{figure=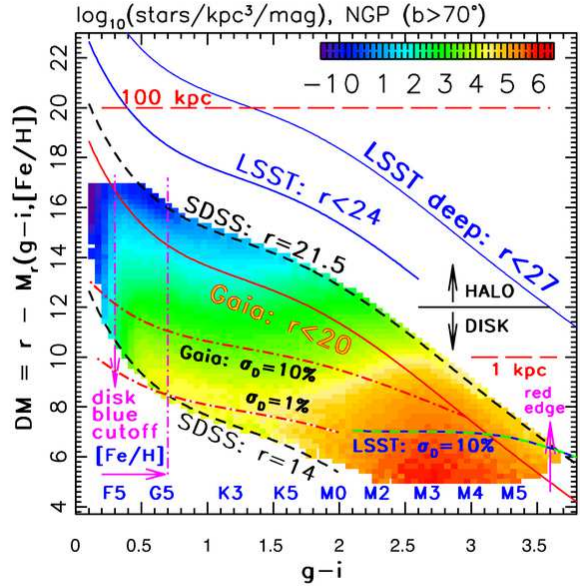,height=25pc}}
\vskip -0.7in
\caption{The volume number density (stars/kpc$^3$/mag, log scale
according to the legend) of main-sequence stars with $14<r<21.5$ and $b>70^\circ$, 
as a function of their distance modulus and $g-i$ color (based on SDSS data). 
The absolute magnitudes are determined using the photometric parallax 
relation from Ivezic08. The MK spectral type is indicated above the $g-i$ axis.
The two vertical arrows mark the turn-off color for disk stars and the red edge of 
the M-dwarf color distribution. The $[Fe/H]$ label shows the color range 
($g-i<0.7$) where the photometric metallicity estimator from Ivezic08 performs
best. The two diagonal dashed lines, marked $r=14$ and $r=21.5$, show 
the apparent magnitude limits for SDSS data. The diagonal solid lines mark 
the apparent magnitude limits for Gaia ($r<20$), LSST's single-epoch data ($r<24$,
10 $\sigma$), and LSST's stacked data ($r<27$, 10 $\sigma$). The dashed
line in the lower right corner marks the distance limits for obtaining 10\% accurate 
trigonometric distances using LSST data. The two dot-dashed lines mark analogous 
limits for obtaining 1\% and 10\% accurate trigonometric distances using Gaia's data.}
\label{GaiaLSSTstarCounts}
\end{figure}

A quantitative comparison of the distance-color coverage for main-sequence
stars by Gaia and LSST is shown in Figure~\ref{GaiaLSSTstarCounts}. For
example, stars just below the main-sequence turn-off, with $M_r=4.5$, will be
detected by Gaia to a distance limit of $\sim$10 kpc ($r<20$), and to
$\sim$100 kpc with LSST's single-epoch data ($r<24.5$). For intrinsically
faint stars, such as late M dwarfs, L/T dwarfs, and white dwarfs, the
deeper limit of LSST will enable detection and characterization of the halo
populations. A star with $M_r=15$ will be detectable to a distance limit of
100 pc with Gaia and $\sim$800 pc with LSST, hence the LSST samples will be
about 100 times larger. In addition, for a substantial fraction of red
stars with $r>20$, LSST will provide trigonometric parallax measurements
accurate to better than 10\%. Hence, despite the unprecedented performance
of Gaia for $r<20$, LSST will enable major discoveries with its deep $r>20$
sky coverage. At the same time, and in addition to its own discoveries,
Gaia will provide excellent astrometric and photometric calibration samples
for LSST. To conclude, ``these are exciting times to study local galaxies''
\citep{Wyse2006}.

\clearpage
\bibliography{IvezicARAArefs} 
\bibliographystyle{Astronomy}
\clearpage

\clearpage

\clearpage

\clearpage

\end{document}